\documentclass[12pt]{article}
\usepackage{embolds}
\usepackage{color}
\usepackage{xfrac}

\usepackage[txfonts]{eriksett}
\usepackage[top=2.5cm,bottom=2.5cm,left=2.5cm,right=2.5cm]{geometry}
\title{Unbiased estimation of the OLS covariance matrix when the errors are clustered}
\author{Tom Boot\footnote{Faculty of Economics and Business, University of Groningen, 9700 AV Groningen, The Netherlands, t.boot@rug.nl.} \and Gianmaria Niccodemi\footnote{Faculty of Humanities, Education and Social Sciences, University of Luxembourg, Esch-sur-Alzette, L-4366 Luxembourg, Luxembourg, gianmaria.niccodemi@uni.lu.} \and Tom Wansbeek\footnote{Faculty of Economics and Business, University of Groningen, 9700 AV Groningen, The Netherlands, t.j.wansbeek@rug.nl.}}

\begin{document}
\maketitle
\begin{abstract}
\noindent
When data are clustered, common practice has become to do OLS and use an estimator of the covariance matrix of the OLS estimator that comes close to unbiasedness. In this paper we derive an estimator that is unbiased when the random-effects model holds. We do the same for two more general structures. We study the usefulness of these estimators against others by simulation, the size of the $t$-test being the criterion. Our findings suggest that the choice of estimator hardly matters when the regressor has the same distribution over the clusters. But when the regressor is a cluster-specific treatment variable, the choice does matter and the unbiased estimator we propose for the random-effects model shows excellent performance, even when the clusters are highly unbalanced.
\end{abstract}
\section{Introduction}
Within-cluster dependence presents a considerable challenge for reliable inference. Even with large data sets, a small number of clusters induces substantial finite sample bias in the estimated variance of the regression coefficients. Several options are available to mitigate this bias. Stata uses a scalar correction to the Liang and Zeger (1986) cluster-robust variance estimator, while Bell and McCaffrey (2002) develop cluster extensions of the MacKinnon and White (1985)\nocite{McKW1985} heteroskedasticity-robust variance estimators. See Cameron and Miller (2015) and MacKinnon, Nielsen, and Webb (2022)\nocite{McKOrrWe22} for recent surveys on the topic. However, with the exception of some special cases, none of these variance adjustments completely eliminates the bias.

In this paper, we develop variance estimators that are unbiased under progressively more complicated dependence structures. Our aim is to investigate whether removing the bias in the variance estimators leads to improved inference, in particular by delivering hypothesis tests with more accurate size control. The key idea underlying the unbiased variance estimator is a cluster extension of the variance estimator by
Hartley, Rao, and Kiefer (1969), which is unbiased under heteroskedasicity. In its original form, this variance estimator has the drawback that it requires inverting a matrix that grows quadratically with the sample size. We show how the underlying structure of this matrix can be exploited to make the computation feasible even with large microeconometric data sets.

With a large number of clusters, test statistics based on cluster-robust variance estimators have a standard normal distribution, see for instance Hansen and Lee (2019)\nocite{HaLe19}. With a small number of clusters, the use of the normal distribution to obtain confidence intervals and critical values can lead to substantial size distortions as discussed in Cameron and Miller (2015), Section VI.D, unless the within-cluster dependence is restricted as in Ibragimov and M\"uller (2016)\nocite{IbMu10}. The use of a $t$-distribution reduces the size distortion, but this requires selecting the appropriate degrees of freedom (d.f.). For our proposed variance estimators, we derive a data-driven estimator for the d.f.\ following the approach based on an independence assumption on the errors as in Bell and McCaffrey (2002) as well as the generalization to a random-effects (RE) structure studied in Imbens and Koles{\'a}r (2016).

We focus on three dependence structures of increasing generality. First, we assume that in each cluster the errors follow the same RE structure. In this case, the covariance structure depends on two (unknown) parameters. Second, we extend this setting by allowing the RE parameters to be cluster dependent, increasing the number of parameters to two times the number of clusters. Finally, we consider a fully unrestricted setting where each cluster has an arbitrary covariance matrix. This captures for example a setting with conditional heteroskedasticity where the covariance matrix depends via an unknown functional form on a set of continuous regressors. Our approach can be readily adapted to a panel data setting.

For each of the three dependence structures, we numerically evaluate the size properties of hypothesis tests based on the unbiased variance estimators. We compare their performance with the default Stata option as well as the HC2 variance estimator by Bell and McCaffrey (2002) with d.f.\ as in Imbens and Koles{\'a}r (2016). The model we consider includes a treatment dummy and a continuous variable. For each covariance structure, we vary the number of treated clusters and consider both a balanced design, where each cluster has the same number of observations, as well as an unbalanced design.

Under the specification where the RE covariance structure is the same across clusters, we find that the corresponding unbiased variance estimator performs remarkably well. Even with only a single treated cluster, hypothesis tests provide accurate size control on both the treatment dummy and the continuous variable. When the number of observations differs between clusters, we find that the d.f.\ calculated under the more general RE assumption improve substantially over those calculated under independence assumptions. In a more general setting where the RE structure is cluster dependent, we find that using the corresponding variance estimator improves over the benchmarks particularly when the design is unbalanced.  Finally, we consider a setting where there is conditional heteroskedasticity that depends on the continuous variable. The most general unbiased variance estimator continues to control size in this set-up.

After these simulations with fully artificial data we compare methods using real-life data with an artificial element added. That is, we estimate a wage equation on the basis of U.S. data, clustered by state. To the real-life data we added an artificial state-wide policy dummy variable. We study the size of an hypothesis test on the effect of this dummy variable by sampling subsets of states either at random or based on their number of observations.

The paper is organized as follows. In Section \ref{basics} we start by deriving the general form of unbiased estimators for error covariance matrices with a linear structure. We then specify this for clusters in   Section \ref{cluster}. We first consider in Section \ref{cluster1} a simple structure with just two parameters, one for the overall error and one for the within-cluster error. In Section \ref{cluster2} we generalize this and make these parameters specific per cluster. In Section \ref{cluster3} we generalize this one more step and allow all covariances within clusters to vary freely. We proceed to compare the performance of the various unbiased variance estimators, first by simulation and then through an application to real-life data. Our performance measure is the size of the $t$-test. The d.f.\ of the $t$-tests play an important role, and in Section \ref{degreesoffreedom} we discuss how we set them. Section \ref{design} describes the set-up of the simulations, while the results are presented in Section \ref{size}. The results for the real-life data are given in Section \ref{real}. Section \ref{conclusion} concludes.

\section{Unbiased variance estimation}
\label{basics}
We consider linear regression $\by=\bX\bbeta+\bepsi$, with $\bX$ exogenous of order $n\times k$. We follow the usual notation $\bM\equiv\bI_n-\bX(\bX'\bX)^{-1}\bX'$ and $\bP\equiv\bX(\bX'\bX)^{-1}\bX'$. The errors are distributed according to $\bepsi\sim(\bzero,\bSigma)$ and we consider the case where $\bSigma$ is linear in parameters,
\[
\mbox{vec}\bSigma=\bD\bpi,
\]
with $\bpi$ of order $r\times 1$ and the design matrix $\bD$ of order $n^2\times r$. We are interested in unbiased estimation of the covariance matrix $\bV$ of the OLS estimator $\hat\bbeta$ of $\bbeta$,
\[
\bV=(\bX'\bX)^{-1}\bX'\bSigma\bX(\bX'\bX)^{-1}.
\]
With
\[
\bR'\equiv\left((\bX'\bX)^{-1}\bX'\otimes(\bX'\bX)^{-1}\bX'\right)\bD,
\]
we have in stacked form, which is more convenient for our analysis,
\begin{eqnarray*}
\bv&\equiv&\mbox{vec}\bV\\
&=&\left((\bX'\bX)^{-1}\bX'\otimes(\bX'\bX)^{-1}\bX'\right)\mbox{vec}\bSigma\\
&=&\bR'\bpi.
\end{eqnarray*}
We base our estimator on a function of the residuals $\hat{\bepsi}\equiv\bM\bepsi$ that is aligned with the structure of $\bSigma$. We hence project the squared residuals on the space spanned by $\bD$, so we use $\bD(\bD'\bD)^{-1}\bD'(\hat{\bepsi}\otimes\hat{\bepsi})$, leading to the estimator
\begin{eqnarray}
\nonumber
\tilde\bv&=&\left((\bX'\bX)^{-1}\bX'\otimes(\bX'\bX)^{-1}\bX'\right)\bD(\bD'\bD)^{-1}\bD'(\hat{\bepsi}\otimes\hat{\bepsi})\\
\label{v0}
&=&\bR'(\bD'\bD)^{-1}\bD'(\hat{\bepsi}\otimes\hat{\bepsi}).
\end{eqnarray}
However, this estimator is biased; with $\E\left(\bD'(\hat{\bepsi}\otimes\hat{\bepsi})\right)=\bD'(\bM\otimes\bM)\bD\bpi$ there holds
\[
\E(\tilde\bv)=\bR'(\bD'\bD)^{-1}[\bD'(\bM\otimes\bM)\bD]\bpi\ne\bR'\bpi=\bv. \]
The bias is easily removed by replacing the term $(\bD'\bD)^{-1}$ by $[\bD'(\bM\otimes\bM)\bD]^{-1}$. For the special case of heteroskedasticity, this idea is due to Hartley, Rao, and Kiefer (1969)\nocite{HaRK69}. The adapted, unbiased estimator of $\bv$ then is
\begin{equation}
\label{hat-v-1}
\hat\bv\equiv\bR'[\bD'(\bM\otimes\bM)\bD]^{-1}\bD'(\hat{\bepsi}\otimes\hat{\bepsi}).
\end{equation}
For computational purposes (\ref{hat-v-1}) is unattractive as the matrix $\bM\otimes\bM$ is huge with large data sets. However, we show below how the simple structure of $\bM$, being the sum of the unit matrix and a matrix of low rank, can be exploited to avoid computational difficulties. A relatively common issue with unbiased estimation of variance components, see for instance Kline, Saggio, and S{\o}lvsten (2020)\nocite{KlSo21}, is that the estimator is not guaranteed to be positive definite. However, corrections that make the estimator positively biased are readily available and avoid overrejection.

Below we will consider three cases, with different design matrices $\bD$. In the third case the number of columns of $\bD$ can be very large. Then we can use an adapted version of (\ref{hat-v-1}). Let
\begin{eqnarray*}
\bA&\equiv&\bD'\bD-\bD'(\bI_n\otimes\bP)\bD-\bD'(\bP\otimes\bI_n)\bD\\
\bW&\equiv&\bX'\bX\otimes\bX'\bX\\
\bF&\equiv&\bD'(\bX\otimes\bX).
\end{eqnarray*}
Then
\begin{eqnarray*}
\bR'&=&\bW^{-1}\bF'\\
\bD'(\bP\otimes\bP)\bD&=&\bF\bW^{-1}\bF'\\
\bD'(\bM\otimes\bM)\bD&=&\bD'\bD-\bD'(\bI_n\otimes\bP)\bD-\bD'(\bP\otimes\bI_n)\bD+\bD'(\bP\otimes\bP)\bD\\
&=&\bA+\bF\bW^{-1}\bF'
\end{eqnarray*}
Since
\[
(\bW+\bF'\bA^{-1}\bF)\bW^{-1}\bF'=\bF'\bA^{-1}(\bA+\bF\bW^{-1}\bF')
\]
there holds
\[
\bW^{-1}\bF'(\bA+\bF\bW^{-1}\bF')^{-1}=(\bW+\bF'\bA^{-1}\bF)^{-1}\bF'\bA^{-1}.
\]
Substitution in (\ref{hat-v-1}) yields
\begin{eqnarray}
\nonumber
\hat\bv&=&\bW^{-1}\bF'(\bA+\bF\bW^{-1}\bF')^{-1}\bD'(\hat{\bepsi}\otimes\hat{\bepsi})\\
\label{hat-v-2}
&=&(\bW+\bF'\bA^{-1}\bF)^{-1}\bF'\bA^{-1}\bD'(\hat{\bepsi}\otimes\hat{\bepsi}).
\end{eqnarray}
This expression still contains the inverse of the matrix $\bA$, which has the same number of columns as $\bD$. It will appear, though, that $\bA^{-1}$ occurs only in the form $\bF'\bA^{-1}$, which appears to have a simple expression in this case.

We now turn to the cluster structure. We denote the number of clusters by $C$ and index them by $c=1,\ldots,C$. Cluster $c$ has $n_c$ observations, so $\sum_cn_c=n$. We let
\begin{eqnarray*}
\ddot{n}&\equiv&\textstyle\sum_cn_c^2\\
\bDelta_n&\equiv&\mbox{diag}\;n_c.
\end{eqnarray*}
Let $\biota_c$ an $n_c$-vector of ones. With a slight abuse of notation we will write $\bI_c$ for $\bI_{n_c}$ and let
\begin{equation}
\label{B}
\bG_c\equiv\left(\begin{array}{c}\bzero\\ \vdots\\ \bI_c\\ \vdots\\ \bzero\end{array}\right)\qquad
\bb_c\equiv\left(\begin{array}{c}\bzero\\ \vdots\\ \biota_c\\ \vdots\\ \bzero\end{array}\right)\qquad
\bB\equiv(\bb_1,\ldots,\bb_c,\ldots,\bb_C).
\end{equation}
The regressors for cluster $c$ are collected in $\bX_c\equiv\bG_c'\bX$ and their sum over the cluster in the row vector $\tilde{\bx}_c'\equiv\bb_c'\bX$. The $\tilde{\bx}_c'$s are collected in the $C\times k$ matrix $\tilde{\bX}\equiv\bB'\bX$. Likewise, $\hat{\bepsi}_c\equiv\bG_c'\hat{\bepsi}$ and $\tilde{\hat{\bepsi}}_c\equiv\bb_c'\hat{\bepsi}$ so $\tilde{\hat{\bepsi}}=\bB'\hat{\bepsi}$.

Below we will frequently perform matrix operations using
\begin{eqnarray*}
\mbox{vec}(\bA\bB\bC)&=&(\bC'\otimes\bA)\mbox{vec}\bB\\
\mbox{tr}(\bA\bB\bC\bD)&=&\mbox{vec}(\bA')'(\bD'\otimes\bB)\mbox{vec}\bC,
\end{eqnarray*}
for conformable generic $\bA,\bB,\bC$ and $\bD$. A piece of notation that is useful in the third case that we will study is the Kronecker product with a dot on top. With $\be_c$ be the $c$th unit vector, we write
\[
\textstyle\sum_c\be_c'\;\dot\otimes\;\bA_c=\left(\bA_1,\ldots,\bA_C\right)
\]
for matrices $\bA_1,\ldots,\bA_C$ with the same number of rows but possibly different number of columns. The use of $\dot\otimes$ is as straightforward as the use of $\otimes$.
\section{Application to three forms of clustering}
\label{cluster}
In this section we consider three, increasingly general structures for $\bSigma$ and present the variance estimator (\ref{hat-v-1}) for each case. The derivations are given in Appendix A.

\subsection{Equicorrelated errors}
\label{cluster1}

We first consider the case where the errors are equicorrelated within clusters, so
\[
\bSigma=\sigma^2\bI_n+\tau^2\bB\bB',
\]
with $\bB$ as given in (\ref{B}). Let
\[
\bPsi=\left(\begin{array}{cc}n-k&n-s\\n-s&\ddot{n}-2\breve{s}+\dot{s}\end{array}\right),
\]
with
\begin{eqnarray*}
s&\equiv&\mbox{tr}(\bX'\bX)^{-1}\tilde{\bX}'\tilde{\bX}\\
\dot{s}&\equiv&\mbox{tr}(\bX'\bX)^{-1}\tilde{\bX}'\tilde{\bX}(\bX'\bX)^{-1}\tilde{\bX}'\tilde{\bX}\\
\breve{s}&\equiv&\mbox{tr}(\bX'\bX)^{-1}\tilde\bX'\bDelta_n\tilde\bX
\end{eqnarray*}
Then
\begin{equation}
\label{v1}
\hat{\bv}=(\bX'\bX\otimes\bX'\bX)^{-1}\left(\mbox{vec}\;\bX'\bX,\mbox{vec}\;\tilde{\bX}'\tilde{\bX}\right)
\bPsi^{-1}(\hat{\bepsi}'\hat{\bepsi},\tilde{\hat{\bepsi}}'\tilde{\hat{\bepsi}})'
\end{equation}
is an unbiased estimator of $\bv$.

Two remarks are in order here. The first one concerns symmetry. The $k\times k$ covariance matrix $\hat\bV$, obtained by rearranging $\hat\bv$ into a matrix, should be symmetric. The derivation of (\ref{v1}) did not take this requirement into consideration. However, it is easy to show that $\hat\bV$ is symmetric, by employing the commutation matrix $\bK_k$, with properties $\bK_k(\bA\otimes\bB)=(\bB\otimes\bA)\bK_k$ for any $k\times k$ matrices $\bA$ and $\bB$ and  $\bK_k\mbox{vec}\bC=\mbox{vec}\bC$ for any symmetric $k\times k$ matrix $\bC$. Symmetry of $\hat\bV$ is equivalent to $\bK_k\mbox{vec}\hat\bv=\mbox{vec}\hat\bv$. By using $\bK_k=\bK_k^{-1}$ this readily follows. The same holds for the other two variance estimators derived below.

The second remark concerns the role played by the regressors. When they would have been neglected in the derivation, that is, estimating $\bv$ by (\ref{v0}) instead of by (\ref{hat-v-1}), we would have obtained
\begin{equation}
\label{psi}
\bPsi=\left(\begin{array}{cc}n&n\\n&\ddot{n}\end{array}\right)\qquad\mbox{so}\qquad
\bPsi^{-1}=\frac{1}{n(\ddot{n}-n)}\left(\begin{array}{rr}\ddot{n}&-n\\-n&n\end{array}\right).
\end{equation}
We can then write
\[
\hat{\bv}=(\bX'\bX\otimes\bX'\bX)^{-1}\left(\mbox{vec}\;\bX'\bX,\mbox{vec}\;\tilde{\bX}'\tilde{\bX}\right)(\hat{\sigma}^2,\hat{\tau}^2)'
\]
or
\begin{equation}
\label{vsimple}
\hat\bV=(\bX'\bX)^{-1}\bX\hat{\bSigma}'\bX(\bX'\bX)^{-1},
\end{equation}
with $\hat\bSigma=\hat{\sigma}^2\bI_n+\hat{\tau}^2\bB\bB'$, where
\begin{eqnarray}
\label{s}
\hat{\sigma}^2&=&\frac{1}{n}\hat{\bepsi}'\hat{\bepsi}-\hat{\tau}^2\\
\label{t}
\hat{\tau}^2&=&\frac{1}{\ddot{n}-n}(\tilde{\hat{\bepsi}}'\tilde{\hat{\bepsi}}-\hat{\bepsi}'\hat{\bepsi}).
\end{eqnarray}
In this form, $\hat\bSigma$ is the estimator for $\bSigma$ used by Imbens and Koles{\'a}r (2016) in their  d.f.\ derivation, to be discussed below in Section \ref{degreesoffreedom}.

\subsection{Cluster-specific parameters}
\label{cluster2}
We next let $\sigma^2$ and $\tau^2$ vary over clusters, so now
\[
\bSigma=\textstyle\sum_c(\sigma_c^2\bG_c\bG_c'+\tau_c^2\bb_c\bb_c').
\]
Let
\[
\bPhi=\left(\begin{array}{cc}\bDelta_n-2\bDelta_s+\bA&\bDelta_n-2\bDelta_{\tilde{s}}+\bL\\
\bDelta_n-2\bDelta_{\tilde{s}}+\bL'&\bDelta_n^2-2\bDelta_n\bDelta_{\tilde{s}}+\bQ\end{array}\right),
\]
with
\begin{eqnarray*}
\Delta_s&=&\mbox{diag}\;\mbox{tr}(\bX'\bX)^{-1}\bX_c'\bX_c\\
\Delta_{\tilde{s}}&=&\mbox{diag}\;\tilde{\bx}_c'(\bX'\bX)^{-1}\tilde{\bx}_c
\end{eqnarray*}
while $\bA, \bL$ and $\bQ$ are matrices of order $C\times C$ with typical elements
\begin{eqnarray*}
a_{cd}&\equiv&\mbox{tr}(\bX'\bX)^{-1}\bX_c'\bX_c(\bX'\bX)^{-1}\bX_d'\bX_d\\
\ell_{cd}&\equiv&\tilde{\bx}_d'(\bX'\bX)^{-1}\bX_c'\bX_c(\bX'\bX)^{-1}\tilde{\bx}_d\\
q_{cd}&\equiv&\left(\tilde{\bx}_c'(\bX'\bX)^{-1}\tilde{\bx}_d\right)^2.
\end{eqnarray*}
Then
\begin{equation}
\label{v2}
\hat{\bv}=(\bX'\bX\otimes\bX'\bX)^{-1}\textstyle\sum_c\left((\mbox{vec}\bX_c'\bX_c)\be_c',(\tilde{\bx}_c\otimes\tilde{\bx}_c)\be_c'\right)
\bPhi^{-1}\textstyle\sum_c(\be_c\hat{\bepsi}_c'\hat{\bepsi}_c,\be_c \tilde{\hat{\bepsi}}^2_c)'
\end{equation}
is the unbiased estimator for the variance $\bv$. 

Also here it is interesting to consider the result when the regressors are neglected. Then
\[
\bPhi=\left(\begin{array}{cc}\bDelta_n&\bDelta_n\\ \bDelta_n&\bDelta_n^2\end{array}\right).
\]
By permuting rows and columns we can rearrange $\bPhi$ into a block-diagonal matrix with $c$th block equal to
\[
\bPhi_c=\left(\begin{array}{cc}n_c&n_c\\n_c&n_c^2\end{array}\right).
\]
from (\ref{psi}) it is clear that this leads to a generalization of (\ref{vsimple}) to the case of cluster-specific parameters, with obvious adaptations of (\ref{s}) and (\ref{t}).
\subsection{Unrestricted error correlation within clusters}
\label{cluster3}
The third case we consider has errors correlate freely within clusters, in a way that differs over clusters. Thus,
\begin{equation}
\label{un}
\bSigma=\mbox{diag}\;\bLambda_c,
\end{equation}
where the $\bLambda_c$ are $n_c\times n_c$ matrices of parameters. With
\[
\bS_c\equiv\bI_{k^2}-\bI_k\otimes\bX_c'\bX_c(\bX'\bX)^{-1}-\bX_c'\bX_c(\bX'\bX)^{-1}\otimes\bI_k
\]
we now obtain
\begin{equation}
\label{v3}
\hat\bv=\left(\bX'\bX\otimes\bX'\bX+\textstyle\sum_c\bS_c^{-1}(\bX_c'\bX_c\otimes\bX_c'\bX_c)\right)^{-1}\textstyle\sum_c\bS_c^{-1}(\bX_c'\hat{\bepsi}_c\otimes\bX_c'\hat{\bepsi}_c)
\end{equation}
as the unbiased estimator of $\bv$ for this case.

Again it is interesting to consider the version of $\hat\bv$ that neglects the regressors. Rearranged into matrix format, it appears to be
\begin{equation}
\label{v5}
\hat\bV=(\bX'\bX)^{-1}\textstyle\sum_c\bX_c'\hat{\bepsi}_c\hat{\bepsi}_c'\bX_c(\bX'\bX)^{-1}.
\end{equation}
This estimator directly generalizes the White (1980)\nocite{Whit80} for cross-sections to clusters and was introduced in the context of panel data analysis by Liang and Zeger (1986), where it underlies the widely used panel-robust standard errors allowing for both heteroskedasticity and correlation over time, see e.g. Cameron and Trivedi (2005)\nocite{CaTr05}.

When the interest shifts from clustered data to panel data one might like to consider the counterpart of (\ref{un}) that is homogeneous over the observational units. The setting then is the panel data model with $N$ units and $T$ waves, so $n=NT$, and the covariance structure is $\bSigma=\bI_N\otimes \bLambda$, with $\bLambda$ of order $T\times T$. We discuss unbiased estimation for this case in Appendix B.
\section{Degrees of freedom}
\label{degreesoffreedom}
The various expressions for $\hat\bV$ or $\hat\bv$ may be of interest by themselves but their main use will be in inference on one particular regression coefficient, $\beta_\ell$, say. For large $C$, the critical values from a standard normal distribution can be used. However, in practice $C$ is often small, and using a $t$-distribution is to be preferred. For instance, Stata uses a $t(C-1)$-distribution after the command \textsc{regress y x, vce(cluster \textit{clustvar})}.

Bell and McCaffrey (2002) proposed a refinement by making the d.f.\ in the $t$-distribution data-dependent. The idea is as follows. Let $v^2_\ell$ be the variance of the OLS estimator $\hat{\beta}_\ell$ and $\hat{v}^2_\ell$ an estimator of $v^2_\ell$. Let
\[
T=\frac{\hat{\beta}_\ell}{v_\ell}/{\frac{\hat{v}_\ell}{v_l}}.
\]
Under normality of the regression errors, the numerator is $N(0,1)$ when $\beta_\ell=0$. Letting $\hat{v}^2_\ell$ be the usual OLS-based estimator of $v^2_\ell$, the denominator is distributed according to
\begin{equation}
\label{vv}
(n-k)\frac{\hat{v}^2_\ell}{v^2_\ell}\sim\chi^2_{n-k},
\end{equation}
leading to the $t(n-k)$-distribution for $T$. This classical result gets lost when we employ another estimator $\hat{v}^2_\ell$ than the usual one, like one of the cluster-robust estimators discussed in Section \ref{cluster}. The proposal of Bell and McCaffrey (2002) is to stay close to (\ref{vv}), by seting the d.f.\ $d_\ell$ such that
\[
d_\ell\;\frac{\hat{v}^2_\ell}{v^2_\ell}\stackrel{\mbox{\tiny{app}}}{\sim}\chi^2_{d_\ell},
\]
where ``app'' stands for ``approximately'' in the sense that the first two moments of $d_\ell\hat{v}^2_\ell/v^2_\ell$ match those of a $\chi^2$-distribution with $d_\ell$ d.f.\ Using unbiased estimators of the variance as derived in the preceding section proves its usefulness here since then the first moments left and right match. Letting the second moments match means $\mbox{var}(d_\ell\hat{v}^2_\ell/v^2_\ell)=2d_\ell$ or
\begin{equation}
\label{dl}
d_\ell=2\frac{(v_\ell^2)^2}{\mbox{var}(\hat{v}^2_\ell)}.
\end{equation}

Obviously, $d_\ell$ is not known and needs to be estimated.
There are two issues with this. One is that $d_\ell$ may depend on parameters, which have to be estimated. A second issue is that evaluating $v^2_\ell$ and $\mbox{var}(\hat{v}^2_\ell)$ requires the distribution of $\bepsi$. As a practical solution to obtain a reasonable value of $\hat{d}_\ell$, Bell and McCaffrey (2002) propose to take $\bepsi\sim N(\bzero,\sigma^2\bI_n)$ as the ``reference distribution.'' Imbens and Koles{\'a}r (2016) suggested to take the RE model as the reference distribution, $\bepsi\sim N(\bzero,\sigma^2\bI_n+\tau^2\bB\bB)$, with $\bB$ as defined in (\ref{B}). We will now derive expressions for $d_\ell$ for both cases. Given our focus on unbiased estimation, we extend previous results by using an unbiased estimator for $\mbox{var}(\hat{v}^2_\ell)$ and by using an unbiased estimator of any parameter that we meet in $d_\ell$.

So, first following Bell and McCaffrey (2002), we let $\bepsi\sim N(\bzero,\sigma^2\bI_n)$. As $\hat{v}^2_\ell$ is quadratic in $\hat\bepsi$, we can write $\hat v^2_\ell=\hat\bepsi'\bA_\ell\hat\bepsi$ for some symmetric $n\times n$ matrix $\bA_\ell$ whose particular form follows from (\ref{v1}), (\ref{v2}) or (\ref{v3}), depending on the case under consideration. For notational simplicity we will omit the subscript $\ell$ to $\bA$ from now on and denote $\ba\equiv\mbox{vec}\bA$, so
\begin{eqnarray*}
\hat v^2_\ell&=&\hat\bepsi'\bA\hat\bepsi\\
&=&\ba'(\hat\bepsi\otimes\hat\bepsi)\\
&=&\ba'(\bM\otimes\bM)(\bepsi\otimes\bepsi)
\end{eqnarray*}
hence
\begin{eqnarray}
\nonumber
\mbox{var}(\hat{v}^2_\ell)&=&2\sigma^4\ba'(\bM\otimes\bM)\ba\\
\label{amam}
&=&2\sigma^4\mbox{tr}\bA\bM\bA\bM.
\end{eqnarray}
From  (\ref{v1}), (\ref{v2}) and (\ref{v3}), $\bA$ readily appears to be block-diagonal, with $c$th block $\bA_c$ given by
\begin{align*}
\bA_{c}&= r_1\bI_{c} + r_2\biota_{c}\biota_{c}', &(r_1,r_2) =&\; \blf_\ell'(\bX'\bX \otimes \bX'\bX)^{-1}(\mbox{vec}\bX'\bX,\mbox{vec}\tilde{\bX}'\tilde{\bX})\bPsi^{-1}\\
\bA_{c}&= r_{1c}\bI_{c} + r_{2c}\biota_{c}\biota_{c}',  &(\br_{1}',\br_{2}')=&\; \blf_\ell'(\bX'\bX\otimes\bX'\bX)^{-1}\textstyle\sum_c\left((\mbox{vec}\bX_c'\bX_c)\be_c',(\tilde{\bx}_c\otimes\tilde{\bx}_c)\be_c'\right)\bPhi^{-1}\\
\bA_{c}&= \bX_{c}\bQ_{c}\bX_{c}', &(\mbox{vec}\bQ_{c})' =&\; \blf_\ell'\left(\bX'\bX\otimes\bX'\bX+\textstyle\sum_c\bS_c^{-1}(\bX_c'\bX_c\otimes\bX_c'\bX_c)\right)^{-1}\bS_c^{-1},
\end{align*}
respectively, with $\blf_\ell\equiv\be_\ell\otimes\be_\ell$ and $\br_1\equiv(r_{11},\ldots,r_{1C})'$ and likewise for $\br_2$. Since $v_\ell^2=\sigma^2\be_\ell'(\bX'\bX)^{-1}\be_\ell$, we obtain
\begin{equation}
\label{hatd1}
d_\ell=\frac{\left(\be_\ell'(\bX'\bX)^{-1}\be_\ell\right)^2}{\mbox{tr}\bA\bM\bA\bM},
\end{equation}
with
\begin{eqnarray}
\nonumber
\text{tr}\bA\bM\bA\bM &=&
\mbox{tr}\textstyle\sum_{c,d}\bG_c\bA_c\bG_c'(\bI-\bP)\bG_d\bA_d\bG_d'(\bI-\bP)\\
\label{tramam}
&=&\text{tr}\textstyle\sum_{c}\bA_{c}^2-2\text{tr} (\bX'\bX)^{-1}\bX'\bA^2\bX+\text{tr}(\textstyle(\bX'\bX)^{-1}
\bX'\bA\bX)^2.
\end{eqnarray}
Computational gains can be had by exploiting the structure of $\bA_c$. Notice that the expression for $d_\ell$ does not depend on unknown parameters since the factor $\sigma^4$ in the numerator and the denominator cancel out.

Next, following Imbens and Koles{\'a}r (2016), we let $\bSigma = \sigma^2\bI_{n} +\tau^2\bB\bB'$, with $\bB$ as defined in (\ref{B}). Instead of (\ref{amam}) we now have $\mbox{var}(\hat{v}^2_\ell)=2\sigma^4\mbox{tr}\bA\bM\bSigma\bM\bA\bM\bSigma\bM$ , and (\ref{hatd1}) generalizes to
\begin{equation}
\label{hatd2}
d_\ell=\frac{\left(\be_\ell'(\sigma^2(\bX'\bX)^{-1}+\tau^2(\bX'\bX)^{-1}
\tilde{\bX}'\tilde{\bX}(\bX'\bX)^{-1})\be_\ell\right)^2}{\mbox{tr}\bA\bM\bSigma\bM\bA\bM\bSigma\bM}.
\end{equation}
Here, both numerator and denominator depend on the parameters $\sigma^4, \tau^4$ and $\sigma^2\tau^2$, which do not cancel out and hence have to be replaced by estimators. The lengthy expression in the denominator posses another complication. Both complications are addressed in Appendix C.

\section{Simulation design}
\label{design}
We take the simulation design of MacKinnon and Webb (2018)\nocite{McKWe18} as our point of departure. The data generating process includes a treatment dummy and a continuous variable. For $c=1,\ldots,C$ it is
\begin{equation}
\label{eq:MCdgp}
\by_c = \biota_c\alpha + \bd_{c}\beta + \bx_{c}\gamma + \bepsi_{c},
\end{equation}
with $\biota_c$ the intercept, $\bd_{c}$ the treatment dummy equal to 1 in clusters $1,\ldots,C_{1}$, which we will vary from $1$ to $C-1$, and $\bx_c$ the continuous regressor, whose elements are independent $N(0,1)$. The regression errors $\bepsi_c$ within cluster $c$ are normally distributed with their covariance matrix $\bSigma_{c}$ specified below. The errors are independent across clusters. We set the parameters $\alpha=\beta=\gamma=0$, the number of clusters $C=14$, and the total number of observations $n=2800$. The results below are based on 200,000 draws of \eqref{eq:MCdgp}. We draw the continuous variable $\bx_c$ only once.

\paragraph{Error covariance matrix} To generate the data, we consider three increasingly complicated designs for the covariance matrix of the $\bepsi_{c}$.
\begin{enumerate}
\item Homogeneous design as Section~\ref{cluster1},
    \begin{equation}
    \label{design1}
    \bSigma_{c} = \sigma^2\bI_{c} + \tau^2\biota_{c}\biota_{c}'.
    \end{equation}
    with $\sigma^2 = 1$ and $\tau^2 = 0.1$.
\item Restricted heterogeneous design as in Section~\ref{cluster2},
    \begin{equation}
    \label{design2}
    \bSigma_{c} = \sigma_{c}^2\bI_{c} + \tau_{c}^2\biota_{c}\biota_{c}'\qquad
    \sigma_{c}^2 = \exp\left(2\delta\frac{C-c}{C-1}\right)\qquad \tau_{c}^2 = \rho \sigma_{c}^2.
    \end{equation}
    This way of including heterogeneity across clusters is borrowed from MacKinnon and Webb (2018). We set $\rho=0.1$ and $\delta=\mbox{ln}(2)/2$, which means that $\sigma_c^2$ ranges from 1 to 2.
\item Unrestricted heterogeneous design as in Section~\ref{cluster3},
    \begin{equation}
    \label{design3}
    \bSigma_{c} = \sigma^2 \bI_{c} + \tau^2\biota_{c}\biota_{c}' + \text{diag}(\bx_{c})^2/2.
    \end{equation}
    with $\sigma^2$ and $\tau^2$ as in the homogeneous design.
\end{enumerate}

\paragraph{Balance}
 An important design choice is the number of observations per cluster. We first consider a balanced design, where the number of observations per cluster is equal to $n/C=200$, and next an unbalanced design, where the number of observations depends on the cluster index according to
\begin{equation}
\label{eq:imbalanced}
\begin{split}
n_{c} &=\mbox{int}\left(n\frac{\exp(\gamma c/C)}{\sum_c\exp(\gamma c/C)}\right), \quad c = 1,\ldots, C-1,\quad n_C= n-\textstyle\sum_cn_c.
\end{split}
\end{equation}
We set $\gamma = 2$, which implies cluster sizes ranging from 67 to 438 observations.

\paragraph{Variance estimators and reference distributions}
We consider the following methods to obtain $t$-values for the OLS estimate for $\beta$ in \eqref{eq:MCdgp}.
\begin{enumerate}
    \item The first benchmark $t$-values are based on  the cluster extension of White's standard errors due to Liang and Zeger (1986)\nocite{liang1986longitudinal} as already introduced in (\ref{v5}), but with a finite-sample correction as implemented in Stata,
    \[
        \hat{\bV}_{\text{LZ1}} = \frac{C}{C-1}\frac{n-1}{n-k} (\bX'\bX)^{-1}\textstyle\sum_{c}\bX_{c}'\hat{\bepsi}_{c}\hat{\bepsi}_{c}'\bX_{c}(\bX'\bX)^{-1}.
    \]
   Following Stata we compare the resulting $t$-statistic against the critical values of a $t(C-1)$ distribution. We denote this benchmark method by STATA.

    \item The second benchmark $t$-values implement the Liang and Zeger (1986)\nocite{liang1986longitudinal} standard errors with a HC2 correction as in Bell and McCaffrey (2002)\nocite{BeMc02}.
    \[
        \hat{\bV}_{\text{LZ2}} = (\bX'\bX)^{-1}\textstyle\sum_{c}\bX_{c}'(\bI_{c}-\bP_{cc})^{-1/2}\hat{\bepsi}_{c}\hat{\bepsi}_{c}'(\bI_{c}-\bP_{cc})^{-1/2}\bX_{c}(\bX'\bX)^{-1},
    \]
    where $\bP_{cc}=\bX_{c}(\bX'\bX)^{-1}\bX_{c}'$.
    We compare the resulting $t$-statistic against the critical values of a $t(d_{\mbox{\tiny{IK}}})$ distribution, with $d_{\mbox{\tiny{IK}}}$ the d.f.\ suggested by Imbens and Koles{\'a}r (2016)\nocite{imbens2016robust}. We denote this benchmark method by LZIK.

    \item We use the three unbiased variance estimators from Sections \ref{cluster1}-\ref{cluster3}, denoted by UV1, UV2, and UV3, respectively, and compare the resulting $t$-statistics against the critical values of a $t$-distribution for both reference distributions considered (indicated by RV0 and RV1, respectively), so with d.f.\ $d_\ell$ from (\ref{hatd1}) and from (\ref{hatd2}). This yields six cases, UV1(RV0), UV1(RV1), UV2(RV0), UV2(RV1), UV3(RV0), and UV3(RV1).
\end{enumerate}
Notice that LZ2 does not exist when the number of (un)treated clusters is smaller than two, and that UV2($\cdot$), UV3($\cdot$) do not exist when the number of (un)treated clusters is smaller than three. We then set the size to zero.

\section{Simulation results}
\label{size}
\begin{figure}[ht]
    \centering
      \caption{Size of the $t$ test for the treatment dummy, SV1}
      \label{fig:Cov1Dummy}
    \includegraphics[width=\textwidth,trim={1cm 1cm 1cm 0},clip]{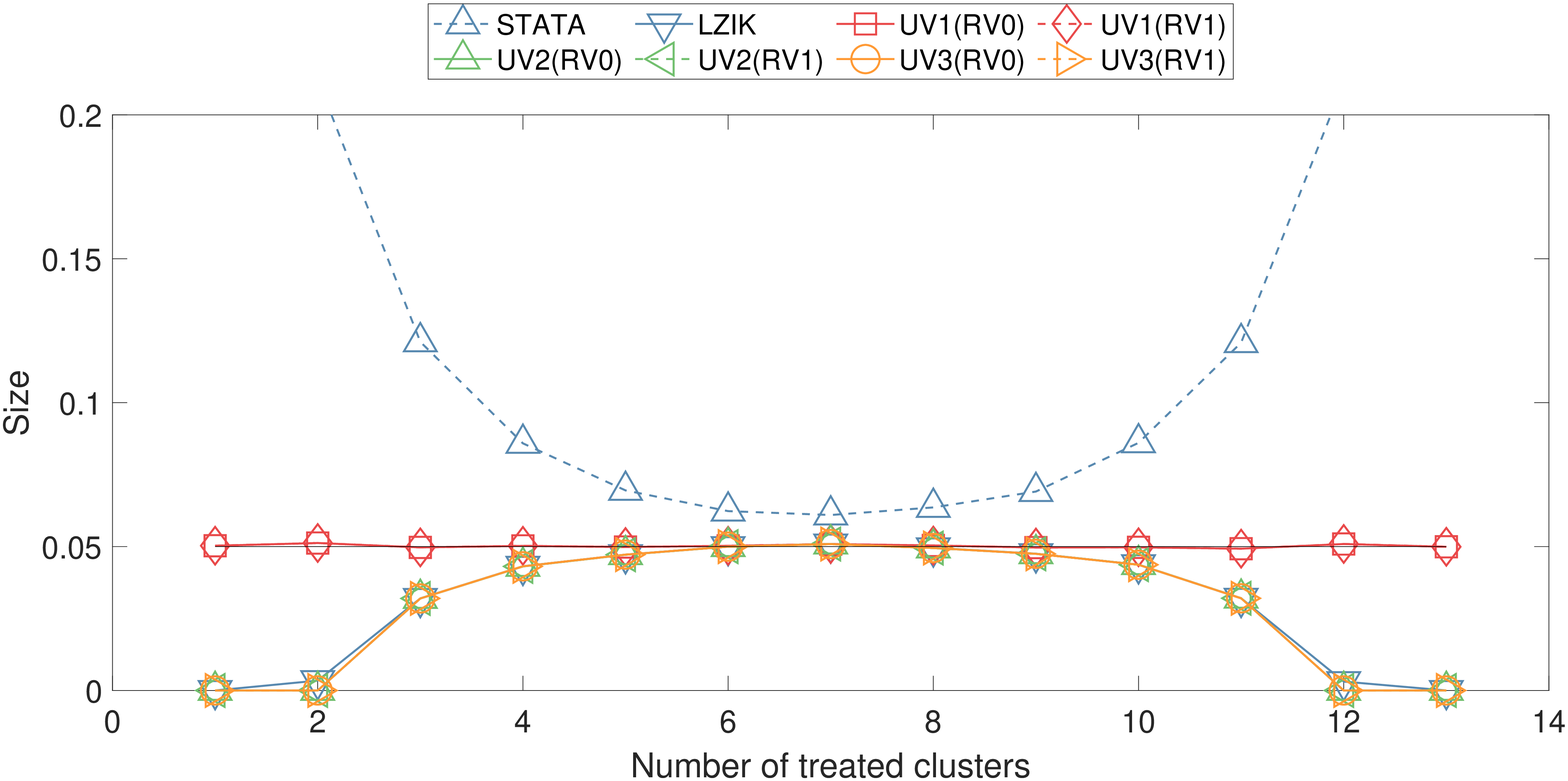}
    \includegraphics[width=\textwidth,trim={1cm 0cm 1cm 0},clip]{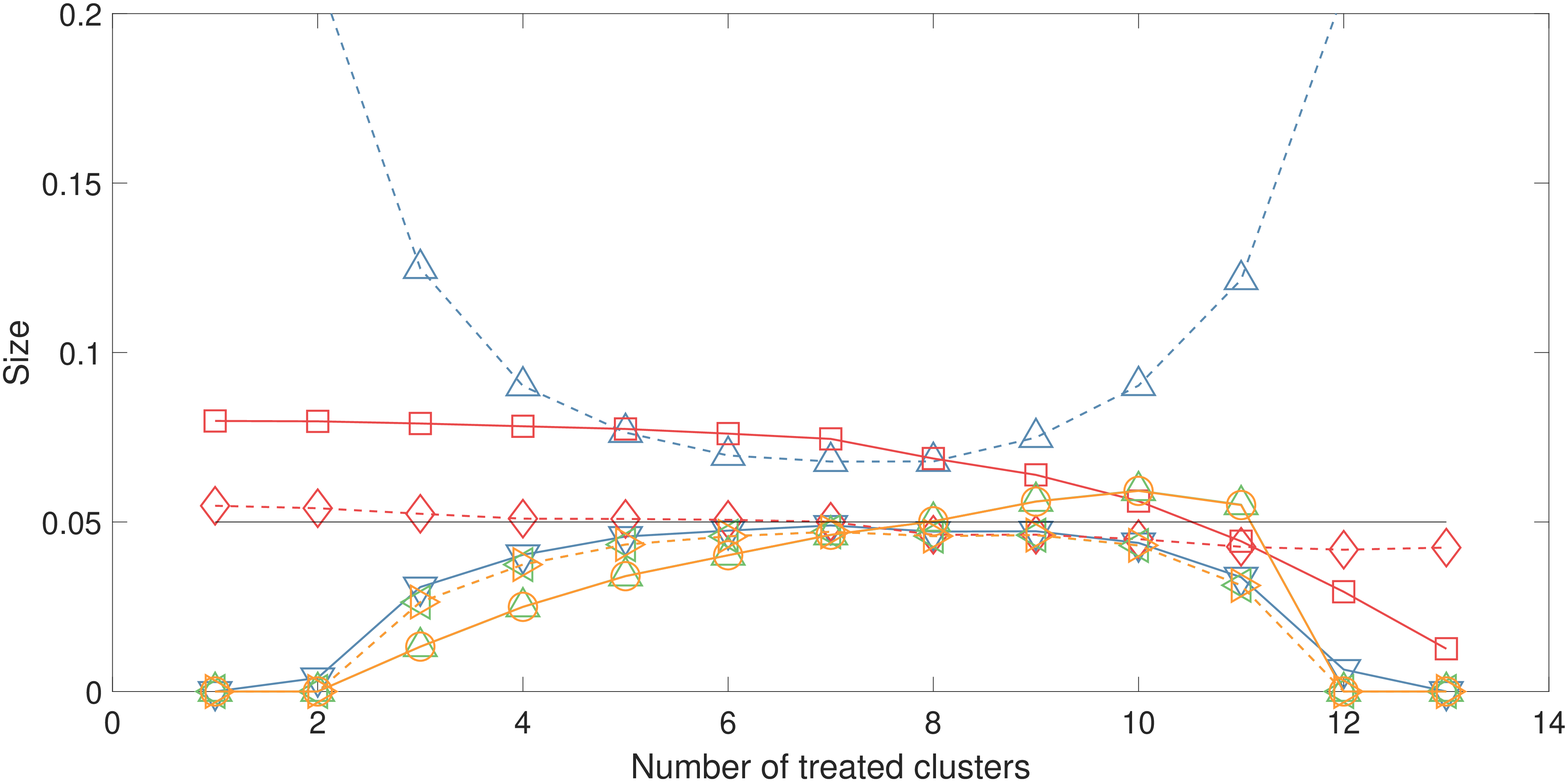}
\end{figure}

\begin{figure}[t]
    \centering
      \caption{Size of the $t$ test for the treatment dummy, SV2}
      \label{fig:Cov2Dummy}
    \includegraphics[width=\textwidth,trim={1cm 1cm 1cm 0},clip]{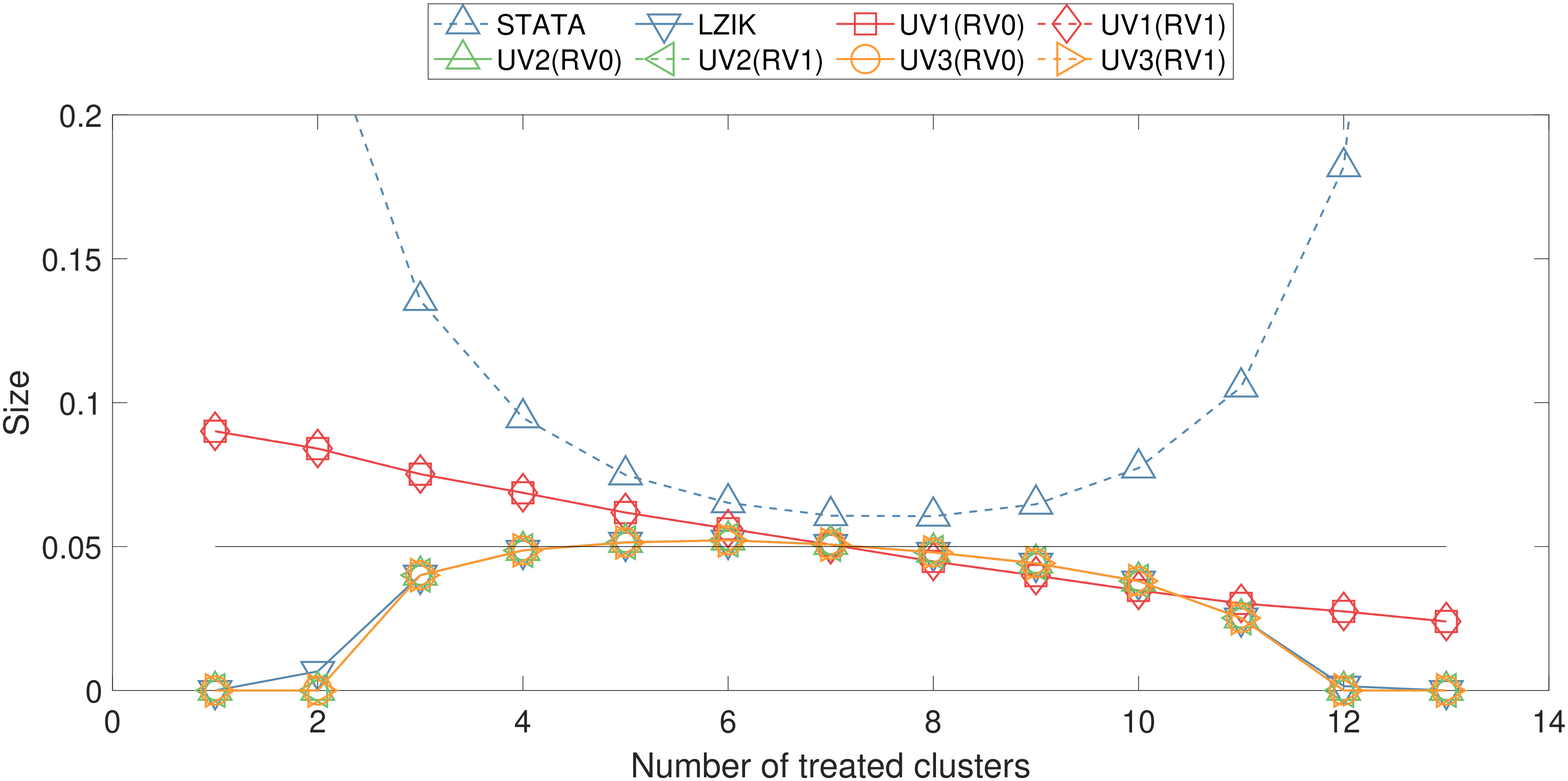}
    \includegraphics[width=\textwidth,trim={1cm 0cm 1cm 0},clip]{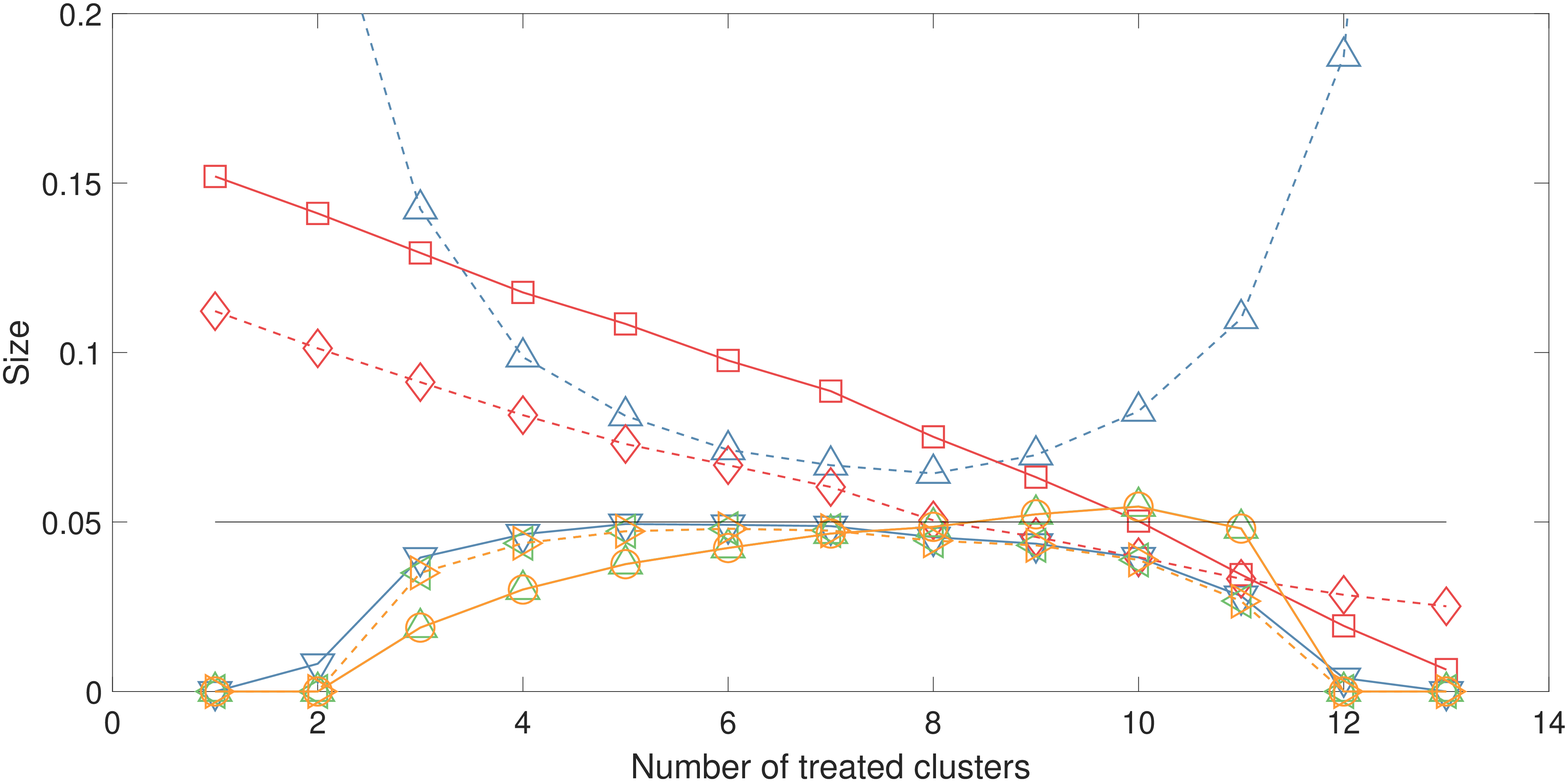}
\end{figure}

\begin{figure}[t]
    \centering
      \caption{Size of the $t$ test for the treatment dummy, SV3}
      \label{fig:Cov3Dummy}
    \includegraphics[width=\textwidth,trim={1cm 1cm 1cm 0},clip]{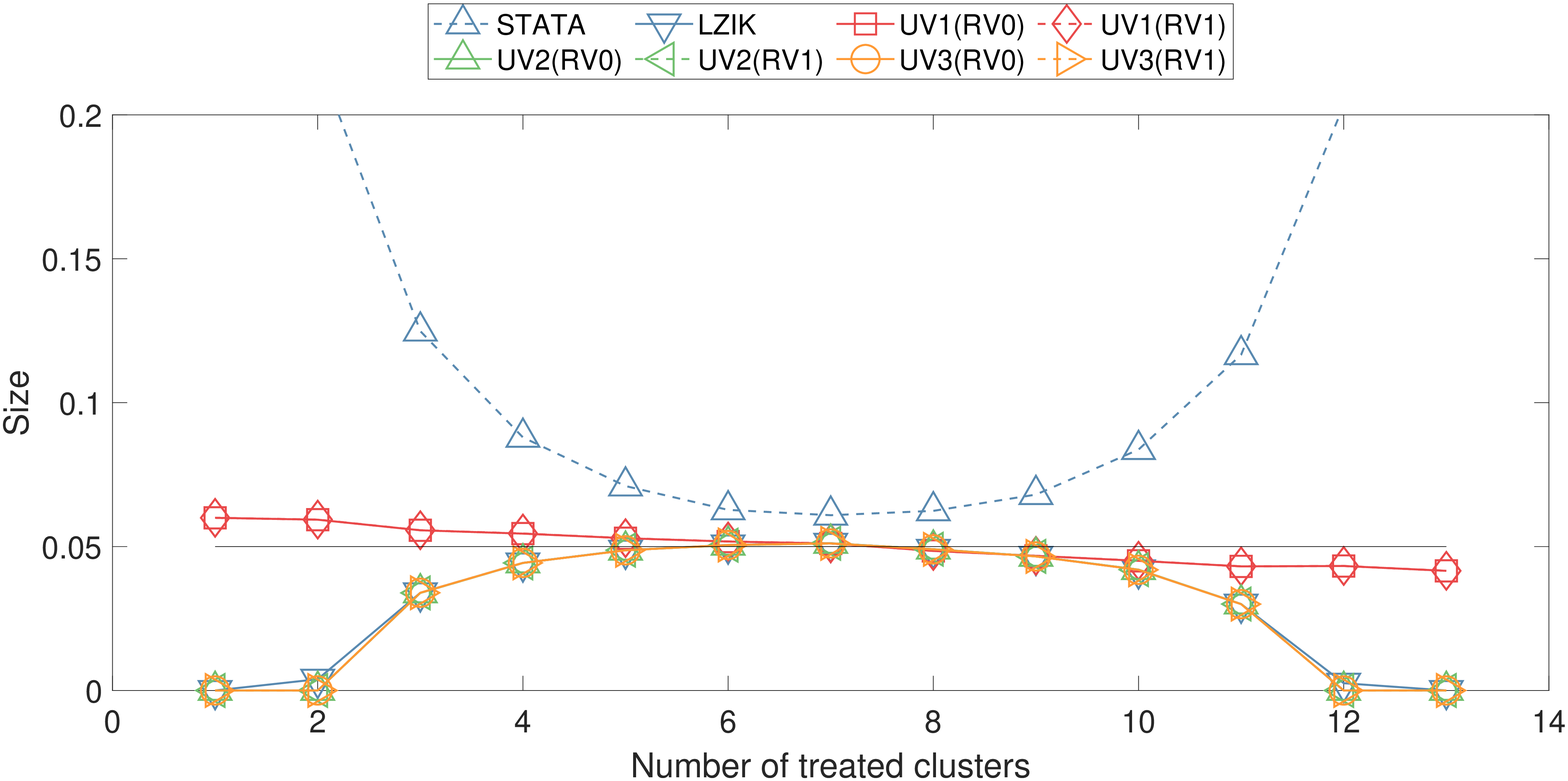}
    \includegraphics[width=\textwidth,trim={1cm 0cm 1cm 0},clip]{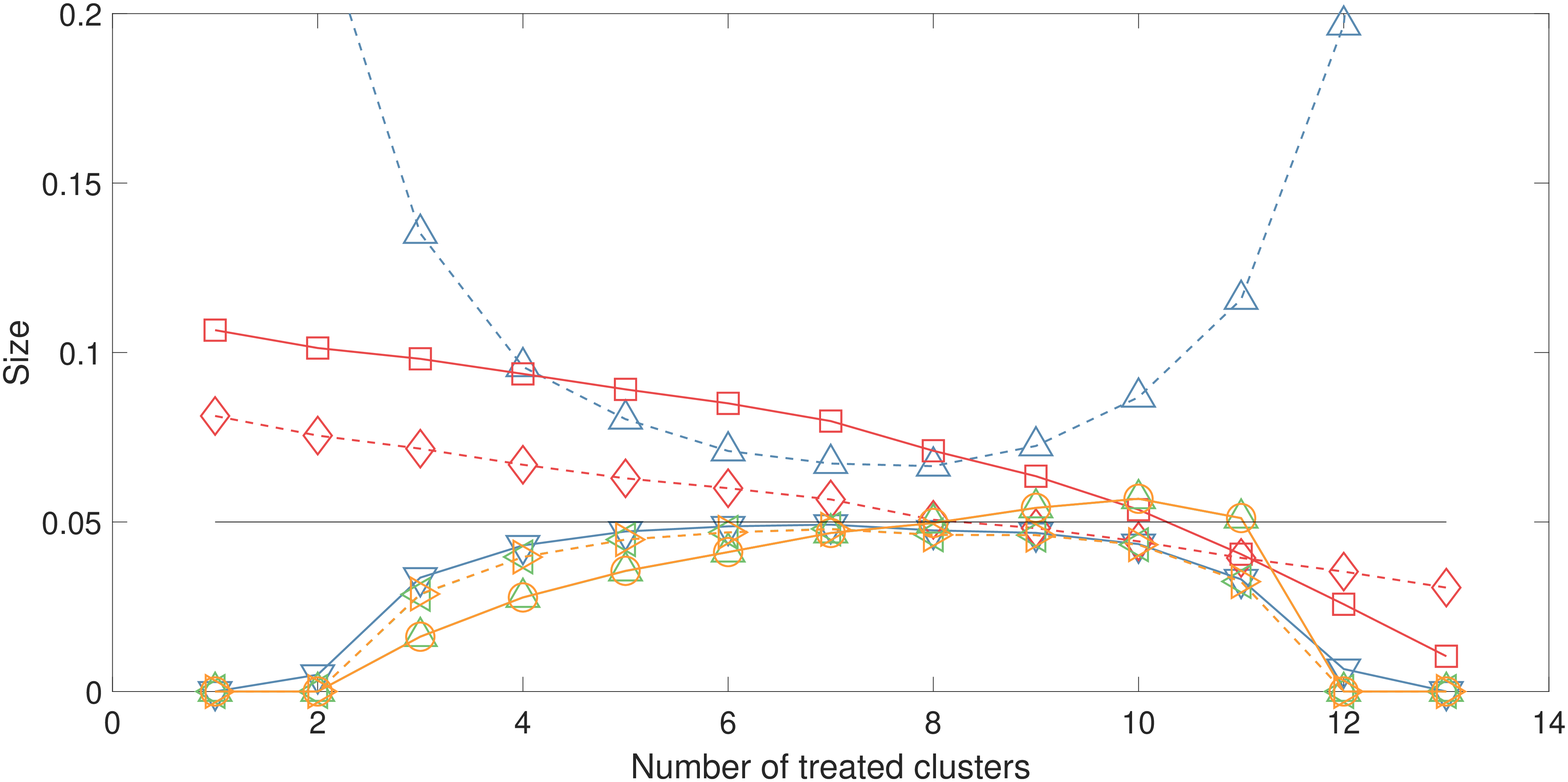}
\end{figure}

The main results of the simulations are presented in Figure \ref{fig:Cov1Dummy}, \ref{fig:Cov2Dummy} and \ref{fig:Cov3Dummy}, based on data simulated with error covariance matrix as in (\ref{design1}), (\ref{design2}) and (\ref{design3}), respectively. They show the size of the $t$-test for $H_0:\beta=0$, with $\beta$ the coefficient of the dummy variable in (\ref{eq:MCdgp}). The number of treated clusters is on the horizontal axis. The upper panel of each figure is for the balanced case and the lower panel for the unbalanced case as described in (\ref{eq:imbalanced}). Each figure shows eight curves, for STATA, LZIK, UV1(RV0), UV1(RV1), UV2(RV0), UV2(RV2), UV3(RV0), and UV3(RV3). Notice that three variances are involved: the reference variance to obtain $d_\ell$; the variance whose unbiased estimator was used; and the variance used in the simulation. For clarity, Table \ref{table1} summarizes.

\begin{table}[h]
\caption{Overview of the variances used}
\label{table1}
\begin{center}
\begin{tabular}{lccc}
\hline\hline
&Reference&Unbiased&Simulation\\
$\bSigma_c$&variance&estimator&variance\\
\hline
$\sigma^2 \bI_c$&RV0&&\\
$\sigma^2 \bI_c+\tau^2\bB\bB'$&RV1&UV1&SV1\\
$\sigma_c^2\bI_c+\tau_c^2\biota_c\biota_c'$&&UV2&SV2\\
$\bLambda_c$&&UV3&SV3\\
\hline\hline
\end{tabular}
\end{center}
\end{table}

The most relevant curves in all three figures are the ones labeled UV1(RV1) in Figure \ref{fig:Cov1Dummy}, upper and lower panels. The homogeneous RE design can be considered the more or less generic case in the clustered-error literature and, as is apparent from Table \ref{table1}, this particular curve is maximally based on this design as it underlies the data generation SV1, the variance estimator UV1, and $d_\ell$ based on RV1.

\paragraph{SV1} Inspecting Figure~\ref{fig:Cov1Dummy} we see, for the balanced design in the upper panel, excellent size control for UV1($\cdot$). This holds even when there is only a single treated cluster. It does not appear to matter whether the d.f.\ are calculated under the more restrictive i.i.d.\ assumption, UV1(RV0), or the RE structure, UV1(RV1). By contrast, UV2($\cdot$), UV3($\cdot$) and LZIK are slightly conservative when we have a small or large number of treated clusters. The STATA variance estimator performs quite poorly, especially when the number of treated clusters is small or large.

Moving to the unbalanced set-up in the lower panel of Figure~\ref{fig:Cov1Dummy}, we see that UV1(RV0) no longer provides accurate size control. However, UV1(RV1), the most relevant case as argued above, still exhibits excellent performance. The additional computational complexity of this approach appears to pay  off. We also see that, unlike in the balanced case, the results for UV2(RV0) and UV3(RV0) differ from the benchmark variance estimator LZIK. The unbiased variance estimators are more conservative for a small number of treated clusters, while becoming slightly oversized for 9-11 treated clusters. UV2(RV1) and UV3(RV1) are again very close to LZIK. The STATA variance estimator again is found not to accurately control size.

\paragraph{SV2} In Figure~\ref{fig:Cov2Dummy} we show the size for $t$-tests based on the various variances estimators under the restricted heterogeneous design where each cluster has its own variance and covariance parameter. This set-up is more general than the homogeneous design in which each cluster has the same variance and covariance parameter. As expected, the performance of UV1(RV0) and UV1(RV1) somewhat deteriorates in this set-up, with size slightly below 0.10 for the case of a single treated cluster and balanced design. The same is observed for an unbalanced design, with the size obtained under UV1(RV1) being just over 0.10.

For UV2 and UV3, under both d.f., and LZIK, we see the test slightly overrejects for a small number of treated clusters. When the number of treated clusters increases, the tests become progressively more conservative. Again, a difference emerges between UV2, UV3 and LZIK in the unbalanced case presented in the lower panel of Figure~\ref{fig:Cov2Dummy}. Here, size control is more accurate for UV2 and UV3 compared to LZIK. Especially UV1(RV0) and UV2(RV0) perform well in this set-up, providing accurate size control up to roughly 8 treated clusters. With more treated clusters, they tend to be conservative, although not as much as LZIK.

\paragraph{SV3} The results for the unrestricted heterogeneous design are nearly identical to those in the homogenous design for the STATA variance, LZIK and UV2 and UV3 under both d.f.\ corrections. For UV1, we find reasonable performance when clusters are balanced. When the clusters are unbalanced, UV1(RV0) becomes oversized for a small number of treated clusters, and undersized when the number of treated clusters is large. The more general d.f.\ correction in UV1(RV1) partly corrects these size distortions.

So far for the test on $\beta$, the coefficient of the cluster-specific dummy variable. We can be much more concise as to $\gamma$, the coefficient of the continuous variable. For SV1 and SV2 the size control is almost perfect. This no longer holds for SV3, where the size is still almost perfect for STATA, LZIK, UV3($\cdot$) but appears to be double the nominal size for $UV1(\cdot$) and UV2($\cdot$); the latter methods are apparently sensitive when the data are generated according to more general scheme SV3.

\begin{figure}[ht]
    \centering
      \caption{Simulations: treatment dummy with homogeneous error covariance matrix. Degrees of freedom.}
      \label{fig:Cov1DummyDOF}
    \includegraphics[width=\textwidth,trim={1cm 1cm 1cm 0},clip]{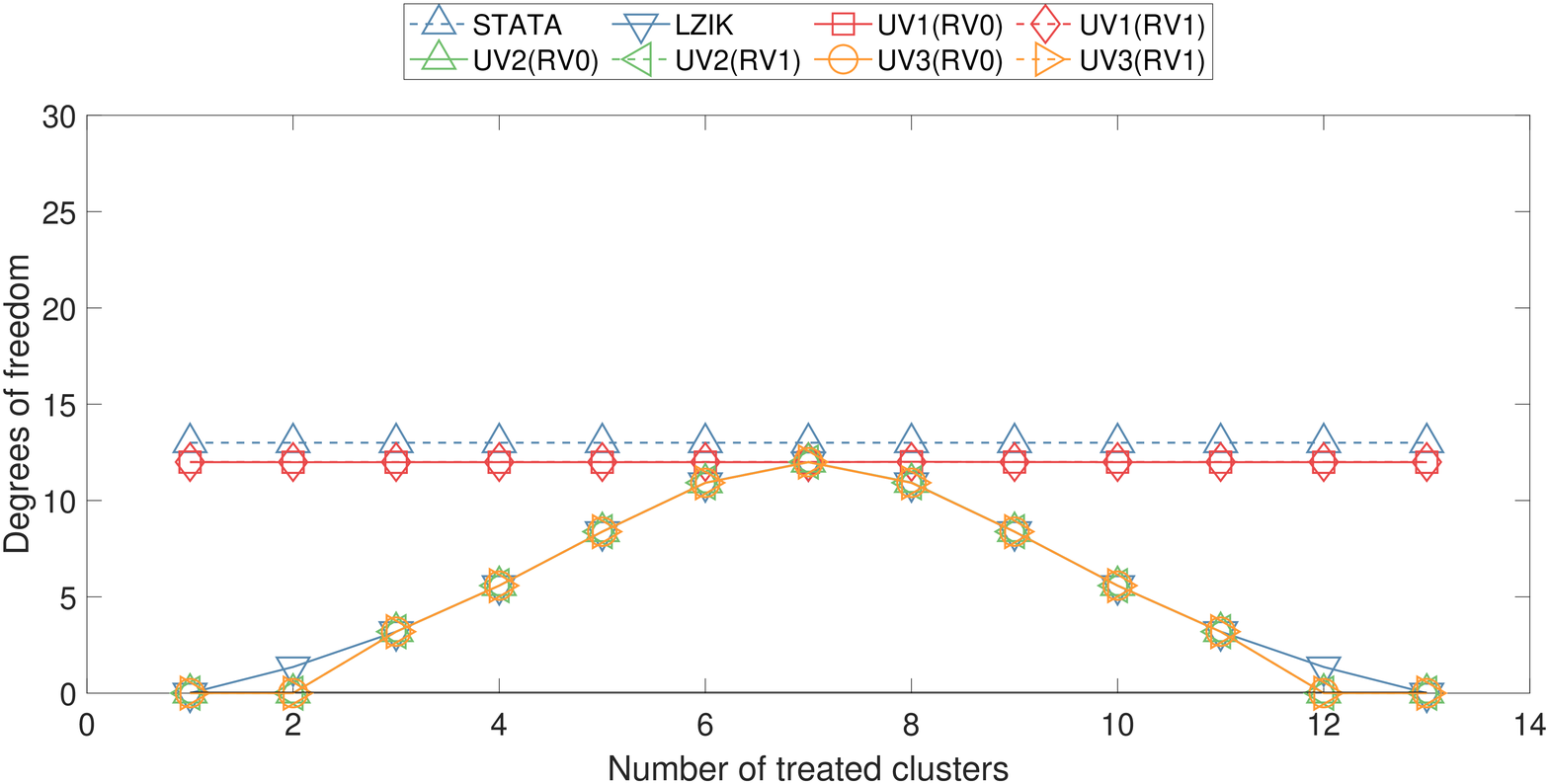}
    \includegraphics[width=\textwidth,trim={1cm 0cm 1cm 0},clip]{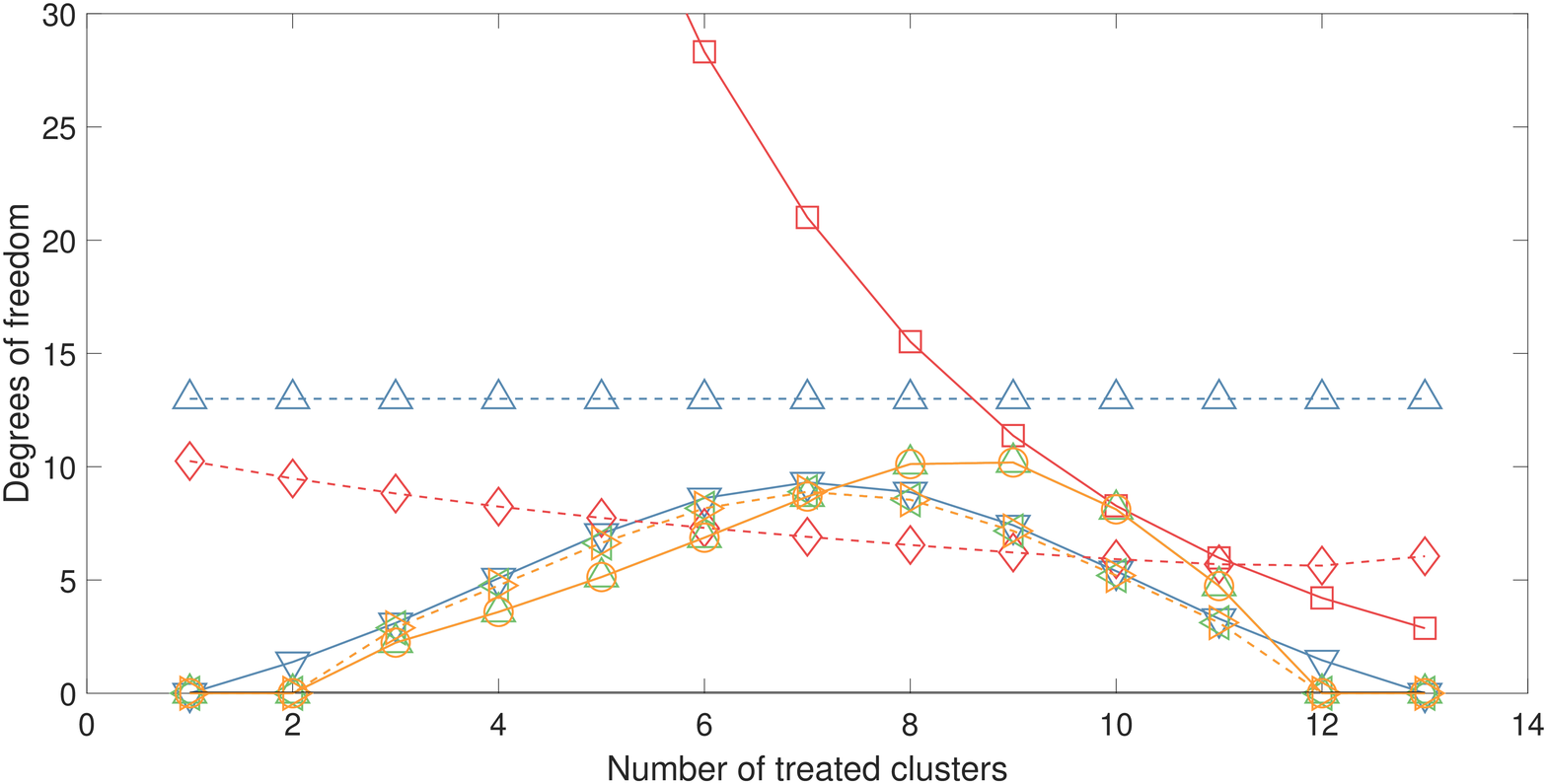}
\end{figure}

\paragraph{Degrees of freedom in SV1} Given the notable differences in performance when using degrees of freedom based on RV0 or RV1, we analyze the degrees of freedom under SV1 in Figure \ref{fig:Cov1DummyDOF}. For a balanced design, we see that the degrees of freedom for UV1 are equal to $C-2$. Donald and Lang (2007)\nocite{DoLa07} show that if the design is balanced and if all regressors are invariant within clusters, the $t$-statistic is $t(C-k)$ distributed, where $k$ is the number of regressors in the model. We can expect the same result to apply here since the continuous variable is uncorrelated with the treatment dummy.

Under a balanced design, the degrees of freedom for the other methods are nearly identical. They are low when the number of treated clusters is low and increase to their maximum value when half of the clusters is treated. This maximum appears to coincide numerically with $C-k$ as well.

When the design is unbalanced, we see a strong deviation from the degrees of freedom under RV0 to those under RV1. This is especially true for UV1 and a small number of treated clusters. For the remaining variance estimators, we see that under RV0 the degrees of freedom are asymmetric in the number of treated clusters, while those under RV1 are symmetric.

\section{Empirical illustration}
\label{real}

To analyze the performance of the unbiased variance estimators in an empirical setting, we consider an application similar to that in Cameron and Miller (2015)\nocite{CaMi15}. We use the Current Population Survey (CPS) 2012 data set that can be obtained from {\tt https://cps.ipums.org/cps/}. The data consist of 51 clusters: the fifty American states and the District of Columbia.  The number of observations in each cluster varies from 519 (Montana) to 5866 (California).

For observation $h$ in cluster $i=1,\dots,C$, we define the model
\begin{equation}
\label{model1}
{\mbox{ln}}(\mbox{\tt wage})_{hi} = \beta_0 + \beta_1 \mbox{\tt educ}_{hi} + \beta_2 \mbox{\tt age}_{hi}+\beta_3 \mbox{\tt age}^{2}_{hi} + \beta_4 \mbox{\tt policy}_{i} + \varepsilon_{hi}.
\end{equation}
Here, $\mbox{\tt policy}$ is a fake policy variable that is randomly
assigned to $C_{1}=1,\ldots,C-1$ sampled clusters and constant within each cluster. Since the policy variable is fake, we expect 5\% rejections across the replications when we test the hypothesis ${\mbox{H}}_0:\beta_4=0$ at the 5\% level.

In line with the simulations in the previous section, we sample a subset of $C=14$ clusters from the 51 available clusters. We consider two different ways of sampling this subset. In the first, we randomly sample clusters with replacement. To test the methods in an unbalanced set-up, we also consider using the $3$ states with the most observations and the $11$ states with the fewest observations. To preserve the relative share of observations in each cluster, we randomly sample with replacement 20\% of the observations within each sampled cluster.

\begin{figure}[t]
    \centering
      \caption{Application: randomly drawn states. Size and degrees of freedom.}
      \label{fig:AppRandom}
    \includegraphics[width=\textwidth,trim={1cm 1cm 1cm 0},clip]{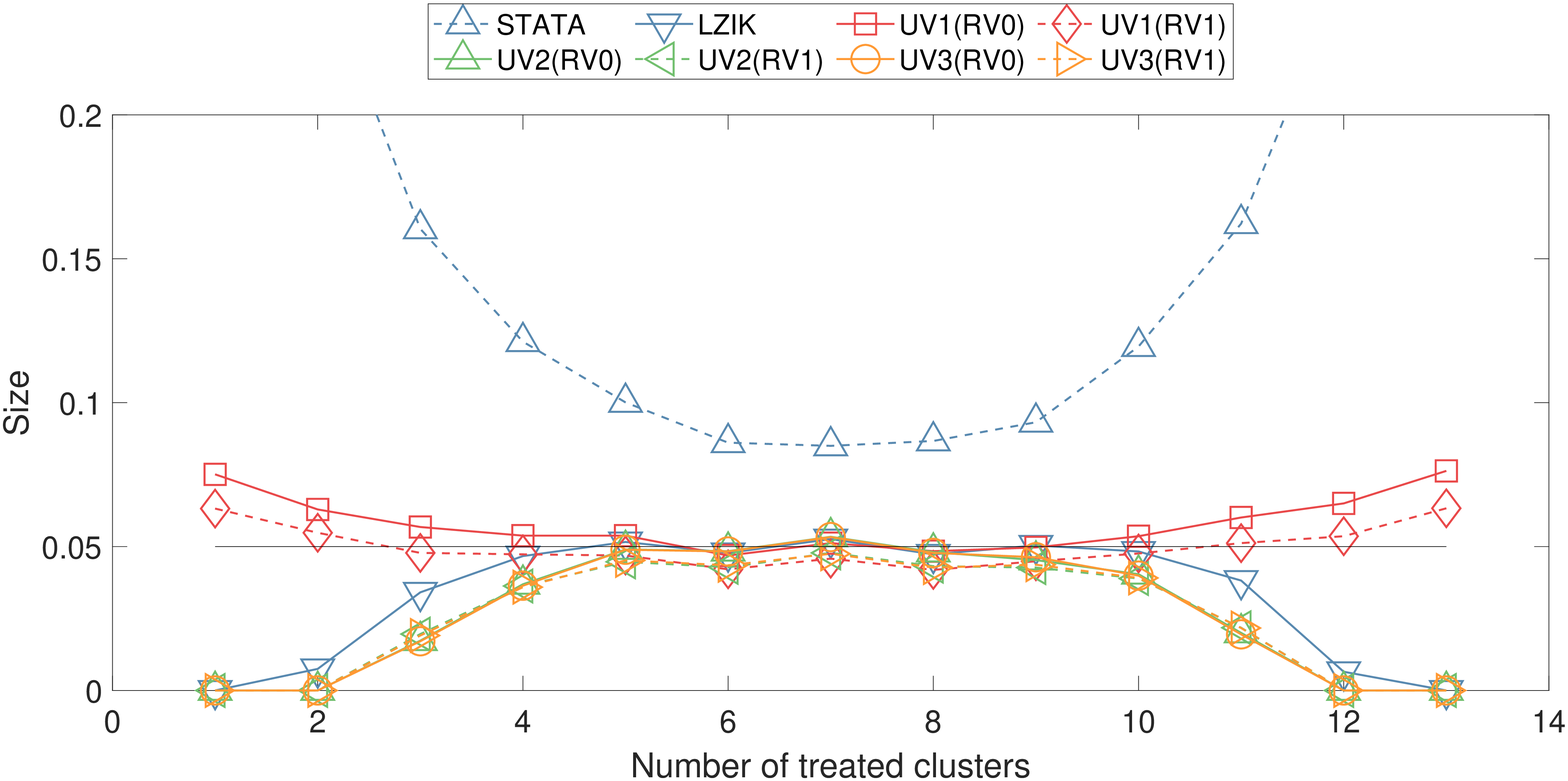}
      \includegraphics[width=\textwidth,trim={1cm 0cm 1cm 0},clip]{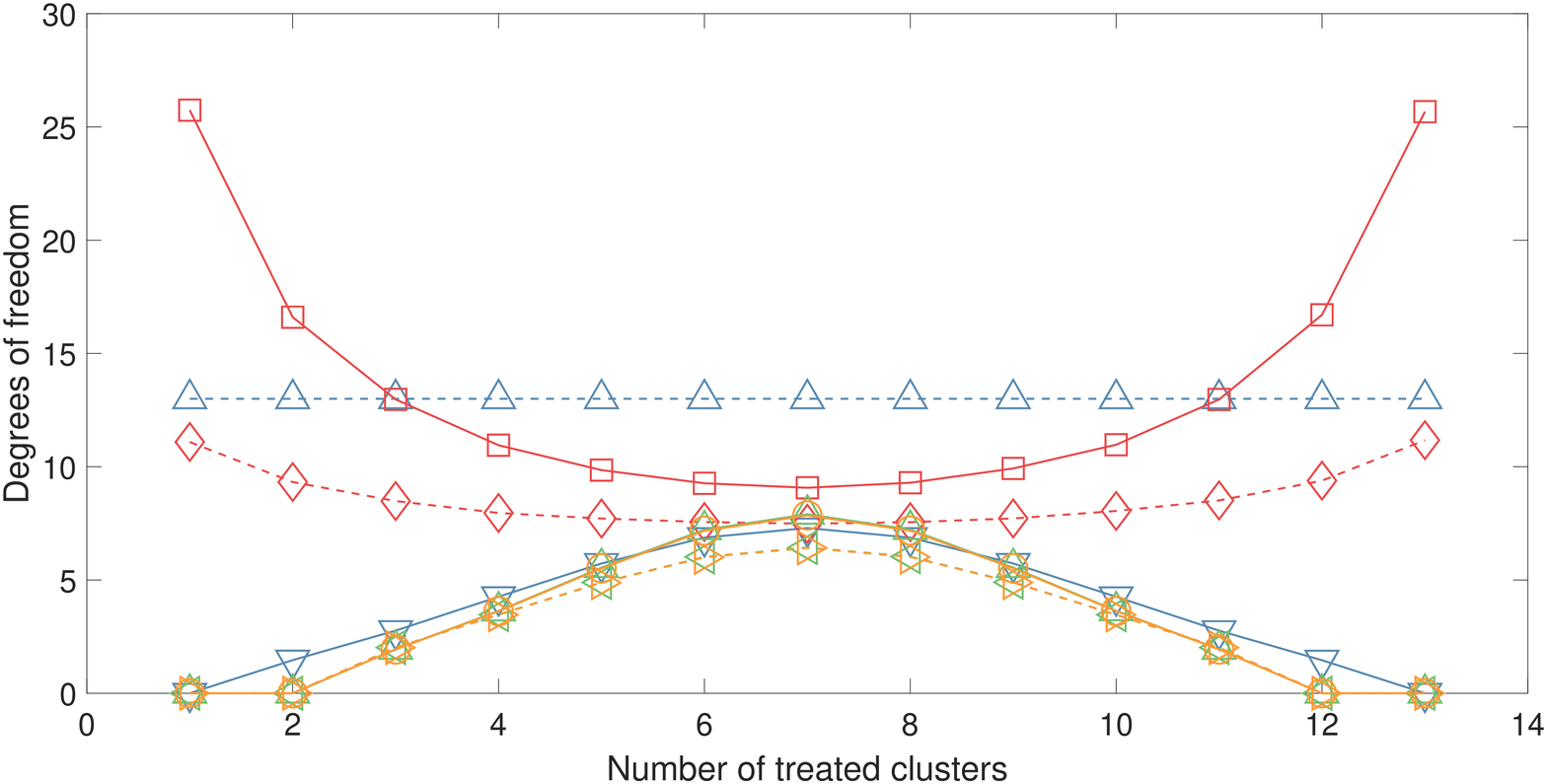}
\end{figure}

\begin{figure}[t]
    \centering
      \caption{Application: ``3-11". Size and degrees of freedom.}
      \label{fig:App311}
    \includegraphics[width=\textwidth,trim={1cm 1cm 1cm 0},clip]{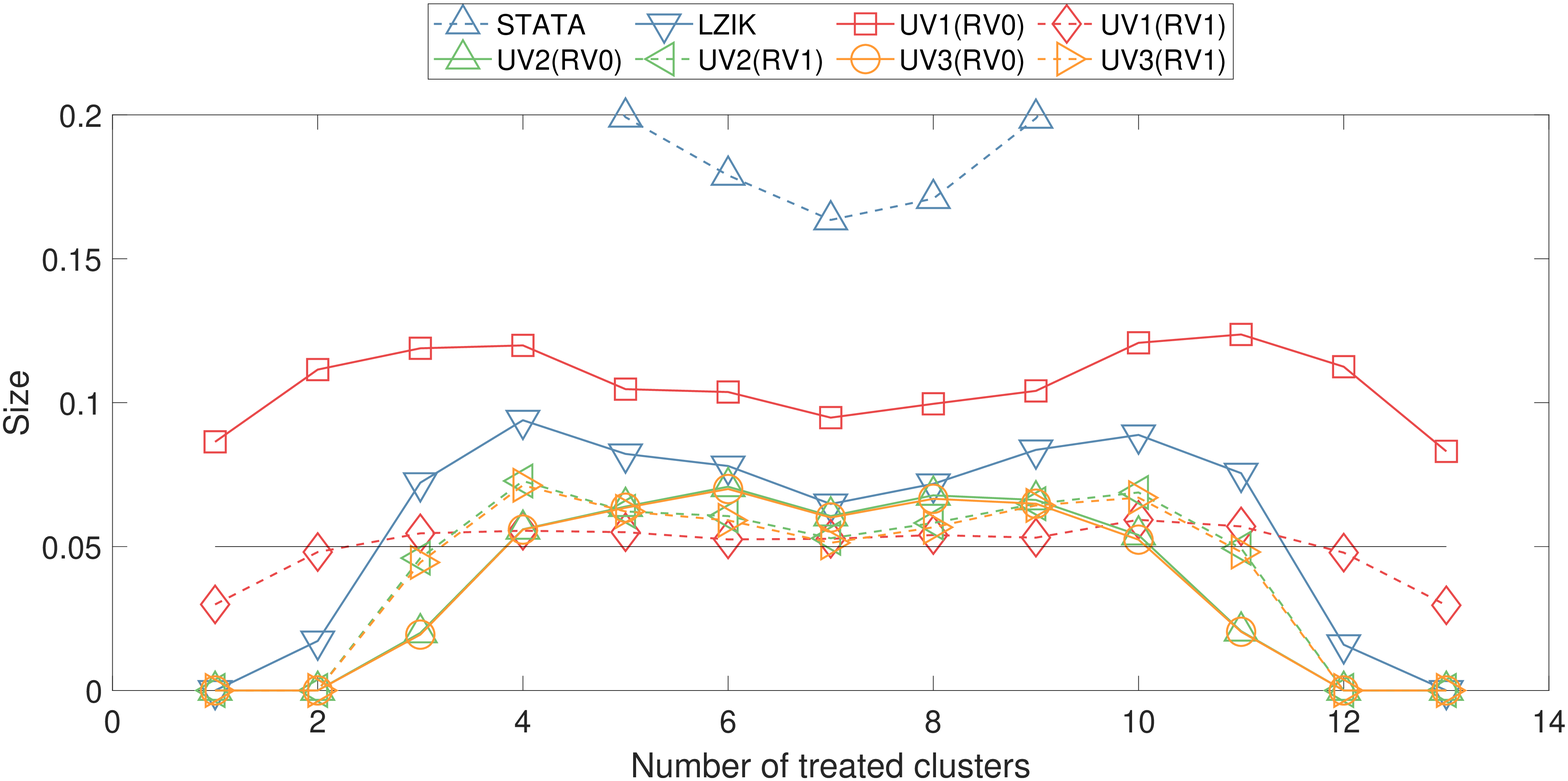}
      \includegraphics[width=\textwidth,trim={1cm 0cm 1cm 0},clip]{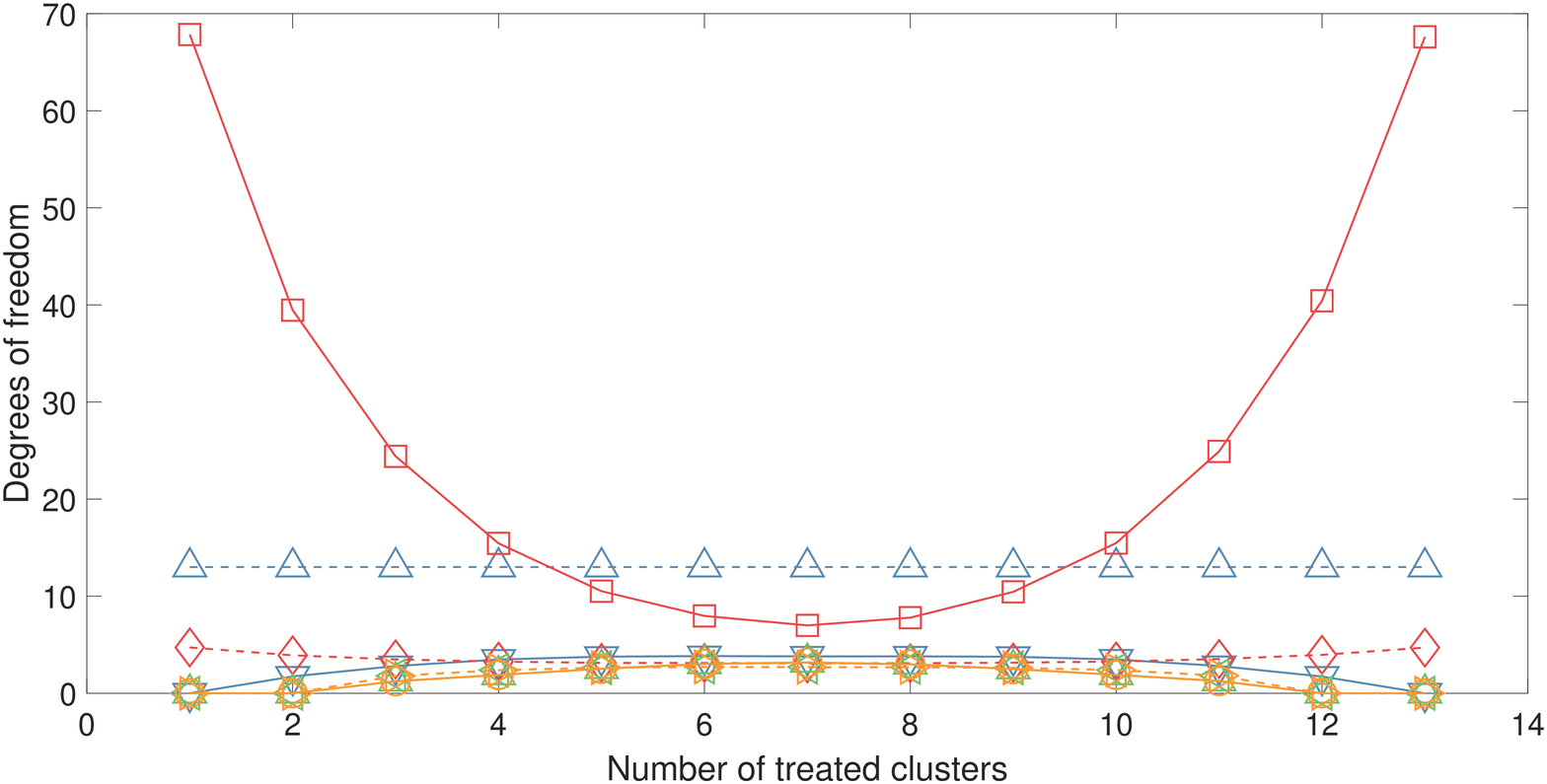}
\end{figure}

Figures \ref{fig:AppRandom}--\ref{fig:App311} show the empirical size (upper panel) and the degrees of freedom (lower panel) averaged over 10,000 replications for the four different designs. The $x$-axis again depicts the number of treated clusters.

In line with the Monte Carlo results from the previous section, we see that the Stata variance estimator with $C-1$ degrees of freedom is severely oversized. In contrast, we find remarkably good size control for UV1(RV1) across the designs. The  degrees of freedom drop considerably when moving from RV0 to RV1. This shows that the use of RV1 is of empirical relevance, especially in the settings with higher imbalance and a small number of (un)treated clusters. The LZIK variance estimator also performs well, although it is oversized in the highly unbalanced ``3-11" setting. There the unbiased variance matrix estimators control size more accurately.

\section{Concluding remarks}
\label{conclusion}
The point of departure in this paper has been to drive unbiased estimators of the covariance matrix of the OLS estimator when the data are clustered. We considered three cases, the leading one being the RE model. This led to our main research question, which is to assess the performance of these estimators in the $t$-test for a particular regression coefficient, both among each other and vis-\`a-vis two oft-used alternatives.

We addressed this question by simulation, in a regression model with a two regressors, one being continuous and distributed equally in all clusters, while the other regressor represented a cluster-specific treatment dummy. The main finding of the simulation study was the excellent behavior of the $t$-test based on the unbiased estimator for the RE model, for the case that the data actually have been generated according to this model and the degrees of freedom have been based on it. So the three variances that play a role are aligned. This result holds for the coefficient of the cluster-specific dummy variable; there is hardly a noticeable difference in performance for between the other variance estimators underlying the $t$-test.

A next step is to see if this excellent behavior also shows up in the case where the three variances are still aligned but now pertain to the more flexible RE model where the two error-components parameters differ over clusters. While by itself this is eminently doable, the question arises to test this heterogeneous RE structure against the homogeneous one. An obvious starting point is the score test context proposed by Breusch and Pagan (1980)\nocite{BrPa80}. Deriving the relevant expression is straightforward but deriving the (limiting) distribution of the test statistic is not since the number of parameters grows with the number of clusters.

The results in the paper on the quality of unbiased estimators in the $t$-test is based on simulation only. We are not aware of any theory that might help giving these results a theoretical basis. There is certainly a research challenge here.
\bibliographystyle{plain-e}
\bibliography{SchmidtEE}

\newpage

\section*{Appendix A: Derivation of the unbiased variance estimators}

\subsection*{Equicorrelated errors}
In this section we consider the case where the errors are equicorrelated within clusters, so
\[
\bSigma=\sigma^2\bI_n+\tau^2\bB\bB',
\]
hence the design matrix for this case is
\[
\bD=(\mbox{vec}\;\bI_n,\mbox{vec}\;\bB\bB').
\]
Let
\begin{eqnarray*}
s&\equiv&\mbox{tr}(\bX'\bX)^{-1}\tilde{\bX}'\tilde{\bX}\\
\dot{s}&\equiv&\mbox{tr}(\bX'\bX)^{-1}\tilde{\bX}'\tilde{\bX}(\bX'\bX)^{-1}\tilde{\bX}'\tilde{\bX}\\
\breve{s}&\equiv&\mbox{tr}(\bX'\bX)^{-1}\tilde\bX'\bDelta_n\tilde\bX.
\end{eqnarray*}
Then
\begin{eqnarray*}
\bD'\bD&=&\left(\begin{array}{cc}
\mbox{tr}\bI_n&\mbox{tr}\bB'\bB\\
\mbox{tr}\bB'\bB&\mbox{tr}(\bB'\bB)^2
\end{array}\right)\\
&=&\left(\begin{array}{cc}n&n\\n&\ddot{n}
\end{array}\right)\\
\bD'(\bI_n\otimes\bP)\bD&=&\left(\begin{array}{cc}
\mbox{tr}\bP&\mbox{tr}\bB'\bP\bB\\
\mbox{tr}\bB'\bP\bB&\mbox{tr}\bB'\bB\bB'\bP\bB
\end{array}\right)\\
&=&\left(\begin{array}{cc}k&s\\s&\breve{s}\end{array}\right)\\
\bD'(\bP\otimes\bP)\bD&=&\left(\begin{array}{cc}
\mbox{tr}\bP&\mbox{tr}\bB'\bP\bB\\
\mbox{tr}\bB'\bP\bB&\mbox{tr}(\bB'\bP\bB)^2
\end{array}\right)\\
&=&\left(\begin{array}{cc}k&s\\s&\dot{s}
\end{array}\right).
\end{eqnarray*}
So
\begin{eqnarray*}
\bPsi&\equiv&\bD'(\bM\otimes\bM)\bD\\
&=&\left(\begin{array}{cc}n-k&n-s\\n-s&\ddot{n}-2\breve{s}+\dot{s}\end{array}\right).
\end{eqnarray*}
So for the current case (\ref{hat-v-1}) becomes
\begin{eqnarray*}
\hat{\bv}&=&\bR'[\bD'(\bM\otimes\bM)\bD]^{-1}\bD'(\hat{\bepsi}\otimes\hat{\bepsi})\\
&=&(\bX'\bX\otimes\bX'\bX)^{-1}(\bX\otimes\bX)'
\left(\mbox{vec}\;\bI_n,\mbox{vec}\;\bB\bB'\right)
\bPsi^{-1}
\left(\mbox{vec}\;\bI_n,\mbox{vec}\;\bB\bB')\right)'(\hat{\bepsi}\otimes\hat{\bepsi})\\
&=&(\bX'\bX\otimes\bX'\bX)^{-1}\left(\mbox{vec}\;\bX'\bX,\mbox{vec}\;\tilde{\bX}'\tilde{\bX}\right)
\bPsi^{-1}(\hat{\bepsi}'\hat{\bepsi},\tilde{\hat{\bepsi}}'\tilde{\hat{\bepsi}})'.
\end{eqnarray*}

\subsection*{Cluster-specific parameters}
We now let $\sigma^2$ and $\tau^2$ vary over clusters and the parameter vector becomes
\[
\blambda=(\sigma^2_1,\ldots,\sigma^2_C,\tau^2_1,\ldots,\tau^2_C)'.
\]
So now
\begin{eqnarray*}
\bSigma&=&\textstyle\sum_c(\sigma_c^2\bG_c\bG_c'+\tau_c^2\bb_c\bb_c')\\
\bD&=&\textstyle\sum_c(\bg_c\be_c',\bh_c\be_c'),
\end{eqnarray*}
with
\begin{eqnarray*}
\bg_c&\equiv&\mbox{vec}\bG_c\bG_c'\\
\bh_c&\equiv&\bb_c\otimes\bb_c,
\end{eqnarray*}
with properties
\begin{eqnarray*}
\bg_c'\bg_c&=&n_c\\
\bh_c'\bh_c&=&n_c^2\\
\bg_c'\bh_c&=&n_c,
\end{eqnarray*}
for $c=1,\ldots,C$, while $\bg_c'\bc_d=\bh_c'\bh_d=\bc_g'\bh_d=0$ for $d\ne c$, and
\begin{eqnarray*}
(\bX\otimes\bX)'\bg_c&=&\mbox{vec}\bX_c'\bX_c\\
(\bX\otimes\bX)'\bh_c&=&\tilde{\bx}_c\otimes\tilde{\bx}_c\\
(\hat{\bepsi}\otimes\hat{\bepsi})'\bg_c&=&\hat{\bepsi}_c'\hat{\bepsi}_c\\
(\hat{\bepsi}\otimes\hat{\bepsi})'\bh_c&=&\tilde{\hat{\bepsi}}^2_c,
\end{eqnarray*}
with $\bepsi_c$ the residuals of cluster $c$ and $\bar{\tilde{\bepsi}}_c$ their sum over the observations in the cluster, this all for $c=1,\ldots,C$. Further
\begin{eqnarray*}
\bg_c'(\bI_n\otimes\bP)\bg_c&=&
(\mbox{vec}\bG_c\bG_c')'\left(\bI_n\otimes\bX(\bX'\bX)^{-1}\bX'\right)(\mbox{vec}\bG_c\bG_c')\\
&=&\mbox{tr}\left(\bG_c\bG_c'\bX(\bX'\bX)^{-1}\bX'\bG_c\bG_c'\right)\\
&=&\mbox{tr}(\bX'\bX)^{-1}\bX_c'\bX_c\\
&\equiv&s_c\\
\bh_c'(\bI_n\otimes\bP)\bh_c&=&
(\bb_c\otimes\bb_c)'\left(\bI_n\otimes\bX(\bX'\bX)^{-1}\bX'\right)(\bb_c\otimes\bb_c)\\
&=&n_c\tilde{\bx}_c'(\bX'\bX)^{-1}\tilde{\bx}_c\\
&\equiv&n_c\tilde{s}_c\\
\bg_c'(\bI_n\otimes\bP)\bh_c&=&
(\mbox{vec}\bG_c\bG_c')'\left(\bI_n\otimes\bX(\bX'\bX)^{-1}\bX'\right)(\bb_c\otimes\bb_c)\\
&=&\mbox{tr}\left(\bG_c\bG_c'\bX(\bX'\bX)^{-1}\bX'\bb_c\bb_c'\right)\\
&=&\tilde{\bx}_c'(\bX'\bX)^{-1}\tilde{\bx}_c\\
&=&\tilde{s}_c,
\end{eqnarray*}
while it appears directly from the derivations that the terms across clusters are zero. This does not hold for the terms involving $\bP\otimes\bP$. There we have
\begin{eqnarray*}
\bg_c'(\bP\otimes\bP)\bg_d&=&
(\mbox{vec}\bG_c\bG_c')'\left(\bX(\bX'\bX)^{-1}\bX'\otimes\bX(\bX'\bX)^{-1}\bX'\right)(\mbox{vec}\bG_d\bG_d')\\
&=&\mbox{tr}\left(\bG_c\bG_c'\bX(\bX'\bX)^{-1}\bX'\bG_c\bG_c'\bX(\bX'\bX)^{-1}\bX'\right)\\
&=&\mbox{tr}(\bX'\bX)^{-1}\bX_c'\bX_c(\bX'\bX)^{-1}\bX_d'\bX_d\\
&\equiv&a_{cd}\\
\bh_c'(\bP\otimes\bP)\bh_d&=&(\bb_c\otimes\bb_c)'
\left(\bX(\bX'\bX)^{-1}\bX'\otimes\bX(\bX'\bX)^{-1}\bX'\right)(\bb_d\otimes\bb_d)\\
&=&\left(\tilde{\bx}_c'(\bX'\bX)^{-1}\tilde{\bx}_d\right)^2\\
&\equiv&q_{cd}\\
\bg_c'(\bP\otimes\bP)\bh_d&=&
(\mbox{vec}\bG_c\bG_c')'\left(\bX(\bX'\bX)^{-1}\bX'\otimes\bX(\bX'\bX)^{-1}\bX'\right)(\bb_d\otimes\bb_d)\\
&=&\tr\left(\bG_c\bG_c'\bX(\bX'\bX)^{-1}\bX'\bb_c\bb_c\bX(\bX'\bX)^{-1}\bX'\right)'\\
&=&\tilde{\bx}_d'(\bX'\bX)^{-1}\bX_c'\bX_c(\bX'\bX)^{-1}\tilde{\bx}_d\\
&\equiv&\ell_{cd}
\end{eqnarray*}
We let $\bDelta_s$ and $\bDelta_{\tilde{s}}$ be the diagonal matrices containing the $s_c$ and $\tilde{s}_c$ and collect the $a_{cd},\ell_{cd}$ and $q_{cd}$ in the matrices $\bA,\bL$ and $\bQ$, respectively. Then we obtain
\begin{eqnarray*}
\bD'\bD&=&\left(\begin{array}{cc}\bDelta_n&\bDelta_n\\ \bDelta_n&\bDelta_n^2\end{array}\right)\\
\bD'(\bI_n\otimes\bP)\bD&=&
\left(\begin{array}{cc}\bDelta_s&\bDelta_{\tilde{s}}\\ \bDelta_{\tilde{s}}&\bDelta_n\bDelta_{\tilde{s}}\end{array}\right)\\
\bD'(\bP\otimes\bP)\bD&=&\left(\begin{array}{cc}\bA&\bL\\ \bL'&\bQ\end{array}\right).
\end{eqnarray*}
So
\begin{eqnarray*}
\bPhi&=&\bD'(\bM\otimes\bM)\bD\\
&=&\left(\begin{array}{cc}\bDelta_n-2\bDelta_s+\bA&\bDelta_n-2\bDelta_{\tilde{s}}+\bL\\
\bDelta_n-2\bDelta_{\tilde{s}}+\bL'&\bDelta_n^2-2\bDelta_n\bDelta_{\tilde{s}}+\bQ\end{array}\right).
\end{eqnarray*}
Combining the various elements, our unbiased estimator of the covariance matrix of the estimated regression coefficients is
\begin{eqnarray*}
\hat{\bv}&=&\bR'[\bD'(\bM\otimes\bM)\bD]^{-1}\bD'(\hat{\bepsi}\otimes\hat{\bepsi})\\
&=&(\bX'\bX\otimes\bX'\bX)^{-1}(\bX\otimes\bX)'
\textstyle\sum_c(\bg_c\be_c',\bh_c\be_c')\bPhi^{-1}\textstyle\sum_c(\bg_c\be_c',\bh_c\be_c')'
(\hat{\bepsi}\otimes\hat{\bepsi})\\
&=&(\bX'\bX\otimes\bX'\bX)^{-1}\textstyle\sum_c\left((\mbox{vec}\bX_c'\bX_c)\be_c',(\tilde{\bx}_c\otimes\tilde{\bx}_c)\be_c'\right)
\bPhi^{-1}\textstyle\sum_c(\be_c\hat{\bepsi}_c'\hat{\bepsi}_c,\be_c \tilde{\hat{\bepsi}}^2_c)'.
\end{eqnarray*}

\subsection*{Unrestricted error correlation within clusters}
We now consider the case where the errors correlate freely within clusters, in a way that differs over clusters. The structure of $\bSigma$ thus is
\begin{eqnarray*}
\bSigma&=&
\mbox{diag}\;\bLambda_c\\
&=&\textstyle\sum_c\bG_c\bLambda_c\bG_c'.
\end{eqnarray*}
This is a quite general structure, involving many parameters. It may even seem too generous in parameters but it has the merit to encompass all kinds of generalizations of the cluster-specific structure of Section \ref{cluster2} like factor structures.  Since
\[
\mbox{vec}\bSigma=\textstyle\sum_c(\bG_c\otimes\bG_c)\mbox{vec}\bLambda_c,
\]
the design matrix now is, using the $\dot\otimes$ notation introduced at the end of Section \ref{basics},
\begin{eqnarray*}
\bD&=&\left(\bG_1\otimes\bG_1,\ldots,\bG_C\otimes\bG_C\right)\\
&=&\textstyle\sum_c\be_c'\;\dot\otimes\;\bG_c\otimes\bG_c.
\end{eqnarray*}
Then, with
\[
\bP_c\equiv\bX_c(\bX'\bX)^{-1}\bX_c',
\]
we obtain
\begin{eqnarray*}
\bD'\bD&=&\textstyle\sum_c\be_c\be_c'\;\dot\otimes\;\bI_c\otimes\bI_c\\
\bD'(\bI_n\otimes\bP)\bD&=&
\left(\textstyle\sum_c\be_c\;\dot\otimes\;\bG_c'\otimes\bG_c'\right)(\bI_n\otimes\bP)
\left(\textstyle\sum_c\be_c'\;\dot\otimes\;\bG_c\otimes\bG_c\right)\\
&=&\textstyle\sum\be_c\be_c'\;\dot\otimes\;\bI_c\otimes\bP_c\\
\bD'(\bP\otimes\bI_n)\bD&=&\textstyle\sum\be_c\be_c'\;\dot\otimes\;\bP_c\otimes\bI_c.
\end{eqnarray*}
In the previous two cases we had a limited amount of parameters. But now we are faced with a possibly very large number of parameters, so we use (\ref{hat-v-2}) rather than (\ref{hat-v-1}).

Elaborating the expressions for $\bA$ and $\bF$ in (\ref{hat-v-2}) for the current case we get
\begin{eqnarray*}
\bA&=&\bD'\bD-\bD'(\bI_n\otimes\bP)\bD-\bD'(\bP\otimes\bI_n)\bD\\
&=&\textstyle\sum_c\be_c\be_c'\;\dot\otimes\;(\bI_c\otimes\bI_c-\bI_c\otimes\bP_c-\bP_c\otimes\bI_c)\\
&\equiv&\textstyle\sum_c\be_c\be_c'\;\dot\otimes\;\bA_c\\
\bF&=&\bD'(\bX\otimes\bX)\\
&=&\textstyle\sum_c\be_c\;\dot\otimes\;\bX_c\otimes\bX_c\\
&\equiv&\textstyle\sum_c\be_c\;\dot\otimes\;\bF_c,
\end{eqnarray*}
with $\bA_c$ and $\bF_c$ implicitly defined. Then
\begin{eqnarray*}
\bF_c'\bA_c&=&\bX_c'\otimes\bX_c'-\bX_c'\otimes\bX_c'\bX_c(\bX'\bX)^{-1}\bX_c'-\bX_c'\bX_c(\bX'\bX)^{-1}\bX_c'\otimes\bX_c'\\
&=&\left(\bI_{k^2}-\bI_k\otimes\bX_c'\bX_c(\bX'\bX)^{-1}-\bX_c'\bX_c(\bX'\bX)^{-1}\otimes\bI_k\right)\bF_c'\\
&\equiv&\bS_c\bF_c',
\end{eqnarray*}
with $\bS_c$ of order $k^2\times k^2$ implicitly defined, so $\bF_c'\bA_c^{-1}=\bS_c^{-1}\bF_c'$ and
\begin{eqnarray*}
\bF'\bA^{-1}\bF&=&\textstyle\sum_c\bF_c'\bA_c^{-1}\bF_c\\
&=&\textstyle\sum_c\bS_c^{-1}(\bX_c'\bX_c\otimes\bX_c'\bX_c).
\end{eqnarray*}
The final expression from (\ref{hat-v-2}) to be elaborated is
\begin{eqnarray*}
\bF'\bA^{-1}\bD'(\hat{\bepsi}\otimes\hat{\bepsi})&=&
\left(\textstyle\sum_c\be_c'\;\dot\otimes\;\bS_c^{-1}\bF_c'\right)
\left(\textstyle\sum_c\be_c\;\dot\otimes\;\bG_c'\otimes\bG_c'\right)
(\hat{\bepsi}\otimes\hat{\bepsi})\\
&=&\textstyle\sum_c\bS_c^{-1}(\bX_c'\hat{\bepsi}_c\otimes\bX_c'\hat{\bepsi}_c).
\end{eqnarray*}
Then (\ref{hat-v-2}) becomes
\[
\hat\bv=\left(\bX'\bX\otimes\bX'\bX+\textstyle\sum_c\bS_c^{-1}(\bX_c'\bX_c\otimes\bX_c'\bX_c)\right)^{-1}\textstyle\sum_c\bS_c^{-1}(\bX_c'\hat{\bepsi}_c\otimes\bX_c'\hat{\bepsi}_c).
\]

\newpage
\section*{Appendix B: Panel data}
Here we continue the discussion at the end of Section \ref{cluster3}, about the panel data model with $N$ units and $T$ waves. The ordering is such that $\bX'=(\bX_1',\ldots,\bX_N')$, with $\bX_i$ of order $T\times k$, for $i=1,\ldots,N$; $\bepsi$ is partitioned likewise. The error structure is $\bSigma=\bI_N\otimes \bLambda$, with $\bLambda$ of order $T\times T$. Since $\sum_i\be_i\be_i'=\bI_T$ we can write
\[
\bSigma=\textstyle\sum_i(\be_i\otimes\bI_T)\bLambda(\be_i'\otimes\bI_T)
\]
and hence
\[
\mbox{vec}\bSigma=\textstyle\sum_i(\be_i\otimes\bI_T\otimes\be_i\otimes\bI_T)\mbox{vec}\;\bLambda.
\]
So the design matrix now is $\bD=\textstyle\sum_i\be_i\otimes\bI_T\otimes\be_i\otimes\bI_T$. Properties are $\bD'\bD=N\bI_{T^2}$ and
\begin{eqnarray*}
(\bI_{NT}\otimes\bX)'\bD&=&\textstyle\sum_i(\bI_N\otimes\bI_T\otimes\bX)'(\be_i\otimes\bI_T\otimes\be_i\otimes\bI_T)\\
&=&\textstyle\sum_i\be_i\otimes\bI_T\otimes\bX_i',
\end{eqnarray*}
so
\begin{eqnarray*}
\bD'(\bI_{NT}\otimes\bP)\bD&=&\bD'(\bI_N\otimes\bI_T\otimes\bX)[\bI_N\otimes\bI_T\otimes(\bX'\bX)^{-1}](\bI_N\otimes\bI_T\otimes\bX)'\bD\\
&=&\textstyle\sum_i(\be_i'\otimes\bI_T\otimes\bX_i)[\bI_N\otimes\bI_T\otimes(\bX'\bX)^{-1}](\be_i\otimes\bI_T\otimes\bX_i')\\
&=&\bI_T\otimes\textstyle\sum_i\bX_i(\bX'\bX)^{-1}\bX_i'\\
&\equiv&\bI_T\otimes\textstyle\sum_i\bP_i,
\end{eqnarray*}
with $\bP_i$ (of order $T\times T$) implicitly defined; likewise $\bD'(\bP\otimes\bI_{NT})\bD=\textstyle\sum_i\bP_i\otimes\bI_T$.
So in this case
\begin{eqnarray*}
\bA&=&N\bI_{T^2}-\bI_T\otimes\textstyle\sum_i\bP_i-\textstyle\sum_i\bP_i\otimes\bI_T\\
\bF&=&\bD'(\bX\otimes\bX)\\
&=&\textstyle\sum_i(\be_i\otimes\bI_T\otimes\be_i\otimes\bI_T)'(\bX\otimes\bX)\\
&=&\textstyle\sum_i\bX_i\otimes\bX_i\\
\hat\bq&=&\bD'({\hat{\bepsi}}\otimes{\hat{\bepsi}})\\
&=&\textstyle\sum_i{\hat{\bepsi}}_i\otimes{\hat{\bepsi}}_i.
\end{eqnarray*}
Substituting this in (\ref{hat-v-2}) yields $\hat\bv=\bW^{-1}\bF'\;\mbox{vec}\;\hat\bLambda$, with $\hat\bLambda$ (symmetric of order $T\times T$) defined by
\[
\mbox{vec}\;\hat\bLambda=(\bA+\bF\bW^{-1}\bF')^{-1}\textstyle\sum_i{\hat{\bepsi}}_i\otimes{\hat{\bepsi}}_i.
\]
The only matrix to be inverted is of order $T^2\times T^2$ so no problem in most cases. Hence
\begin{equation}
\label{var-panel}
\hat\bV=(\bX'\bX)^{-1}\left(\textstyle\sum_i\bX_i'\hat\bLambda\bX_i\right)(\bX'\bX)^{-1}.
\end{equation}
The corresponding expression that does not aim for unbiasedness is simply obtained by letting $\hat{\bLambda}=\sum_i{\hat{\bepsi}}_i{\hat{\bepsi}}_i'/N$.

\newpage
\section*{Appendix C: Degrees of freedom with random effects}
In this appendix we elaborate the denominator of (\ref{hatd2}) and derive estimators for the parameters in $\hat{d}_\ell$. We start with the former. First,
\begin{equation}
\label{amsmamsm}
\mbox{tr}\bA\bM\bSigma\bM\bA\bM\bSigma\bM=\sigma^4\mbox{tr}\bA\bM\bA\bM+2\sigma^2\tau^2\mbox{tr}\bB'\bM\bA\bM\bA\bM\bB+\tau^4\mbox{tr}(\bB'\bM\bA\bM\bB)^2.
\end{equation}
The first term at the right-hand side was already elaborated in (\ref{tramam}). As to the second term,
\[
\bA\bM\bA=\textstyle\sum_c\bG_c\bA_c^2\bG_c'-
\textstyle\sum_{c,d}\bG_c\bA_c\bX_c(\bX'\bX)^{-1}\bX_d'\bA_d\bG_d'
\]
so
\begin{eqnarray*}
\mbox{tr}\bB'\bM\bA\bM\bA\bM\bB&=&\mbox{tr}\textstyle\sum_c\bA_c^2\bG_c'\bM\bB\bB'\bM\bG_c\\
&&-\mbox{tr}\textstyle\sum_{c,d}(\bX'\bX)^{-1}\bX_d'\bA_d\bG_d'\bM\bB\bB'\bM\bG_c\bA_c\bX_c.
\end{eqnarray*}
From
\begin{eqnarray*}
\bG_c'\bM\bB&=&\biota_c\be_c'-\bX_c(\bX'\bX)^{-1}\tilde\bX'\\
&\equiv&\biota_c\be_c'-\bL_c
\end{eqnarray*}
we obtain
\[
\bG_c'\bM\bB\bB'\bM\bG_c=\biota_c\biota_c'-\bX_c(\bX'\bX)^{-1}\tilde\bx_c\iota_c'-\biota_c\tilde\bx_c'(\bX'\bX)^{-1}\bX_c'+\bX_c(\bX'\bX)^{-1}\tilde\bX'\tilde\bX(\bX'\bX)^{-1}\bX_c'
\]
and, letting $\bmu_c\equiv(\bX'\bX)^{-1}\bX_c'\bA_c\biota_c$, we have $\mbox{tr}\bB'\bM\bA\bM\bA\bM\bB=T_1+T_2$, with
\begin{eqnarray*}
T_1&=&\textstyle\sum_c\biota_c'\bA_c^2\biota_c-2\textstyle\sum_c\biota_c'\bA_c^2\bX_c(\bX'\bX)^{-1}\tilde\bx_c+\mbox{tr}\textstyle\sum_c(\bX'\bX)^{-1}\tilde\bX'\tilde\bX(\bX'\bX)^{-1}\bX_c'\bA_c^2\bX_c\\
T_2&=&\mbox{tr}\textstyle\sum_{c,d}(\bX'\bX)^{-1}\bX_d'\bA_d(\biota_d\be_d'-\bL_d)
(\be_c\biota_c'-\bL_c')e\bA_c\bX_c\\
&=&\textstyle\sum_c\bmu_c'\bX'\bX\bmu_c-2\textstyle\sum_c\tilde\bx_c'(\bX'\bX)^{-1}\bX'\bA\bX\bmu_c
+\mbox{tr}(\bX'\bX)^{-1}\bX'\bA\bX(\bX'\bX)^{-1}\tilde\bX'\tilde\bX(\bX'\bX)^{-1}\bX'\bA\bX.
\end{eqnarray*}
So far for the second term at the right-hand side of (\ref{amsmamsm}).

As to the third term, let $\lambda_c\equiv\biota_c'\bA_c\biota_c$ and \begin{eqnarray*}
\bB'\bM\bA\bM\bB&=&\textstyle\sum_c
(\bB'-\tilde\bX(\bX'\bX)^{-1}\bX')\bG_c\bA_c\bG_c'(\bB-\bX(\bX'\bX)^{-1}\tilde\bX')\\
&=&\textstyle\sum_c\left(\lambda_c\be_c\be_c'-\tilde\bX\bmu_c\be_c'-\be_c\bmu_c'\tilde\bX'\right)+\tilde\bX\bW\tilde{\bX}'\\
&\equiv&\bS+\tilde\bX\bW\tilde{\bX}'.
\end{eqnarray*}
Then
\begin{eqnarray*}
\mbox{tr}\bS^2&=&\textstyle\sum_c\left(\lambda_c^2-4\be_c'\tilde\bX\bmu_c+2\bmu_c'\tilde\bX'\tilde\bX\bmu_c\right)
+2\textstyle\sum_{c,d}\tilde\bx'\bmu_d\be_d'\tilde\bx'\bmu_c\\
\mbox{tr}\bS\tilde\bX\bW\tilde{\bX}'&=&\textstyle\sum_c\left(\lambda_c\tilde{\bx}_c'\bW\tilde{\bx}_c-2\tilde{\bx}_c'\bW\tilde\bX\bmu_c\right)\\
\mbox{tr}(\tilde\bX\bW\tilde{\bX}')^2&=&\mbox{tr}\left(\bW\tilde{\bX}'\tilde\bX\right)^2.
\end{eqnarray*}
Combining these elements we obtain an expression for $\mbox{tr}(\bB'\bM\bA\bM\bB)^2$.

In the spirit of the ``unbiased'' theme of this paper, we estimate $d_\ell$ in (\ref{dl}) by using unbiased estimators for $\sigma^4, \sigma^2\tau^2$ and $\tau^4$, which we will now derive. With the subscript to $\blm_{ab}$ denoting an expression with $a$ $\bM$s and $b$ $\bB\bB'$s, there holds
\begin{eqnarray*}
\E(\hat\bepsi\ast\hat\bepsi)&=&\bH'\E(\hat\bepsi\otimes\hat\bepsi)\\
&=&\bH'(\bM\otimes\bM)\mbox{vec}(\sigma^2\bI_n+\tau^2\bB\bB')\\
&=&\sigma^2\bH'\mbox{vec}\bM+\tau^2\bH'\mbox{vec}\bM\bB\bB'\bM\\
&\equiv&\sigma^2\blm_{10}+\tau^2\blm_{21}.
\end{eqnarray*}
We additionally have
\begin{eqnarray*}
\E(\hat\bepsi\ast\bB\bB'\hat\bepsi)&=&\bH'\E(\hat\bepsi\otimes\bB\bB'\hat\bepsi)\\
&=&\bH'(\bM\otimes\bB\bB'\bM)\mbox{vec}(\sigma^2\bI_n+\tau^2\bB\bB')\\
&=&\sigma^2\bH'\mbox{vec}\bB\bB'\bM+\tau^2\bH'\mbox{vec}\bB\bB'\bM\bB\bB'\bM\\
&\equiv&\sigma^2\blm_{11}+\tau^2\blm_{22}
\end{eqnarray*}
and
\begin{eqnarray*}
\E(\bB\bB'\hat\bepsi\ast\bB\bB'\hat\bepsi)&=&\bH'\E(\bB\bB'\hat\bepsi\otimes\bB\bB'\hat\bepsi)\\
&=&\bH'(\bB\bB'\bM\otimes\bB\bB'\bM)\mbox{vec}(\sigma^2\bI_n+\tau^2\bB\bB')\\
&=&\sigma^2\bH'\mbox{vec}\bB\bB'\bM\bB\bB'+\tau^2\bH'\mbox{vec}\bB\bB'\bM\   \bB\bB'\bM\bB\bB'\\
&\equiv&\sigma^2\blm_{12}+\tau^2\blm_{23}.
\end{eqnarray*}
Then
\begin{eqnarray*}
\biota_n'\E(\hat\bepsi\ast\hat\bepsi\ast\hat\bepsi\ast\hat\bepsi)
&=&3\biota_n'\left((\sigma^2\blm_{10}+\tau^2\blm_{21})\ast(\sigma^2\blm_{10}+\tau^2\blm_{21})\right)\\
\biota_n'\E(\hat\bepsi\ast\hat\bepsi\ast\bB\bB'\hat\bepsi\ast\bB\bB'\hat\bepsi)
&=&\biota_n'\left((\sigma^2\blm_{10}+\tau^2\blm_{21})\ast(\sigma^2\blm_{12}+\tau^2\blm_{23})\right.\\
&&\left.+2(\sigma^2\blm_{11}+\tau^2\blm_{22})\ast(\sigma^2\blm_{11}+\tau^2\blm_{22})\right)\\
\biota_n'\E(\bB\bB'\hat\bepsi\ast\bB\bB'\hat\bepsi\ast\bB\bB'\hat\bepsi\ast\bB\bB'\hat\bepsi)
&=&3\biota_n'\left((\sigma^2\blm_{12}+\tau^2\blm_{23})\ast(\sigma^2\blm_{12}+\tau^2\blm_{23})\right).
\end{eqnarray*}
Solving the sample counterpart of this system readily leads to unbiased estimators for the three parameters,
\[
\left(\begin{array}{ccc}
3\biota_n'(\blm_{10}\ast\blm_{10})&6\biota_n'(\blm_{10}\ast\blm_{21})&3\biota_n'(\blm_{21}\ast\blm_{21})\\
x&y&z\\
3\biota_n'(\blm_{12}\ast\blm_{12})&6\biota_n'(\blm_{12}\ast\blm_{23})&3\biota_n'(\blm_{23}\ast\blm_{23})
\end{array}\right)
\left(\begin{array}{c}
\widehat{\sigma^4}\\ \widehat{\sigma^2\tau^2}\\ \widehat{\tau^4}
\end{array}\right)=
\left(\begin{array}{c}
\textstyle\sum_i\hat{\eps}_i^4\\
\textstyle\sum_i\hat{\eps}_i^2\tilde{\eps}_i^2\\
\textstyle\sum_i\tilde{\eps}_i^4
\end{array}\right),
\]
with
\begin{eqnarray*}
x&\equiv&\biota_n'(\blm_{10}\ast\blm_{12}+2\blm_{11}\ast\blm_{11})\\
y&\equiv&\biota_n'(\blm_{10}\ast\blm_{23}+\blm_{21}\ast\blm_{12}+4\blm_{22}\ast\blm_{11})\\
z&\equiv&\biota_n'(\blm_{21}\ast\blm_{23}+2\blm_{22}\ast\blm_{22}).
\end{eqnarray*}
Efficient computation can be based on $\bH'\mbox{vec}\bR\bS'=(\bR\ast\bS)\biota_\ell$ for $\bR$ and $\bS$ of order $n\times \ell$.

\newpage
\section*{}

\end{document}